\documentclass[11pt]{article}
\usepackage[utf8]{inputenc}
\usepackage{listings}
\usepackage{color}
\usepackage{graphicx}
\usepackage{longtable}
\usepackage{float}
\usepackage{amsmath}
\usepackage{amssymb}
\usepackage{authblk}
\usepackage{subcaption}

\title{}

\date{}
\begin{document}

\title{Grid-free stochastic simulations of reaction-diffusion processes at cell-cell contacts}
\author{Thorsten Pr\"ustel} 
\author{Martin Meier-Schellersheim} 
\affil{Computational Systems Biology Section\\Laboratory of Immune System Biology\\National Institute of Allergy and Infectious Diseases\\National Institutes of Health, Bethesda, Maryland 20892, USA}
\maketitle
\let\oldthefootnote\thefootnote 
\renewcommand{\thefootnote}{\fnsymbol{footnote}} 
\footnotetext[1]{Email: prustelt@niaid.nih.gov} 
\let\thefootnote\oldthefootnote

\begin{abstract}
Biological cells can exchange messages through soluble molecules or membrane-bound receptors. In particular in the latter case, the interaction is usually located in specific regions of the interacting cells and may depend on or induce local morphological features or reorganizations of the membrane-associated or membrane-proximal biochemistry. Examples are interactions among cells that probe their environment and the surfaces of neighboring cells with dendrite-like protrusions. Previously, we introduced an algorithm capable of creating spatially-resolved, stochastic particle-based computational models of cellular biochemistry. However, its applicability was limited to the case of soluble ligands. Here, building on this earlier approach, we introduce a simulation method that accounts for the unique features that distinguish the sites of cell-cell contact from other regions of the cell membrane. Employing a smooth, grid-free computational representation of the contact sites, we extend the applicability of Green's function-based stochastic simulation algorithms to study intermembrane bimolecular interactions between particles on adjacent curved surfaces. We demonstrate the utility of this computational approach by exploring the effects of contact geometry and local biochemistry for the early phase of T-cell receptor activation. 
\end{abstract}
\section{Introduction}\label{introduction} 
Most signaling processes in biological cells known to us are initiated by the ligation of membrane-bound receptors on the cell surface. The ligands can be soluble, as is typically the case for growth factors, hormones and chemokines. However, receptor-ligand interactions also occur in a cell-cell contact-dependent manner, where ligands are displayed on the surface of adjacent cells. These juxtacrine signaling events play an important role during development, in tissue homeostasis and nervous as well as immune system function \cite{manz2010spatial, belardi2020cell}.  

Previously, we introduced an algorithm capable of incorporating stochastic particle dynamics within and on the surface of arbitrarily shaped morphologies \cite{prustel2020stochastic}. This algorithm is well suited to generate spatially-resolved computational models of cellular signaling processes that originate from receptor interactions with soluble ligands, such as cellular responses to an extracellular gradient of ligands. It employs a well-known implicit surface modeling technique to provide realistic smooth and grid-free computational representations of spatial domains. Additionally, the method uses a Green's function-based approach \cite{prustel2021space} to enhance the efficiency of Brownian Dynamics simulations to study interactions among particles on curved surfaces and with others located in adjacent bulk (volume) regions, without introducing an artificial discretization length scale.

However, this earlier algorithm could not account for the geometry and local biochemistry of cell-cell contact interfaces, because such interfaces entail geometric constraints that have no analogue in soluble ligand mediated cell signaling. Here, we introduce a generalization of the earlier algorithm that overcomes this limitation, thereby permitting stochastic particle-based simulations of signaling processes involving interactions among membrane-bound receptors on adjacent cells.

When intercellular information exchange takes place on cell-cell contact sites, the cells' morphological and topographical aspects potentially play a greater role than in contact-independent cellular communication, because they impose additional constraints on the local biochemistry \cite{delon2000information, manz2010spatial, fernandes2019cell, belardi2020cell}. One instance of such a constraint is the intermembrane gap distance. This distance acts a 'physical' regulator of the local biochemistry by determining whether membrane complexes are close enough to be able to form intercellular bonds at all. In turn, the diffusive motion of these intermembrane bound complexes may slow down substantially or even come to a halt due to the coupling between the cells \cite{chen2021trapping}. At the same time, a close contact site may passively exclude those membrane proteins whose ectodomains are too large to fit into the space separating the interacting cells. Thus, this interplay of various length scales--gap distance and ectodomain size -- may locally alter the spatial organization of membrane complexes, restrict molecular interactions and thereby strongly impact the local biochemistry and information exchange between the two cells \cite{delon2000information, cordoba2013large, chang2016initiation, belardi2020cell}.

A prominent case in point is the mechanism underlying the Kinetic Segregation model (KS) asserting that the large ectodomain of the CD45 phosphatase leads to its passive exclusion from the sites of TCR engagement \cite{davis2006kinetic, chang2016initiation}, thereby creating regions with diminished phosphatase activity that thus foster the phosphorylation of TCR-proximal signaling components.
\begin{figure}
\includegraphics[scale=0.6]{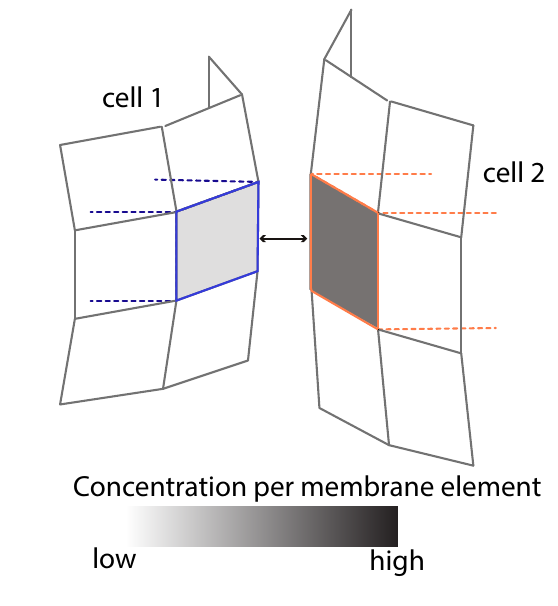}
\includegraphics[scale=0.6]{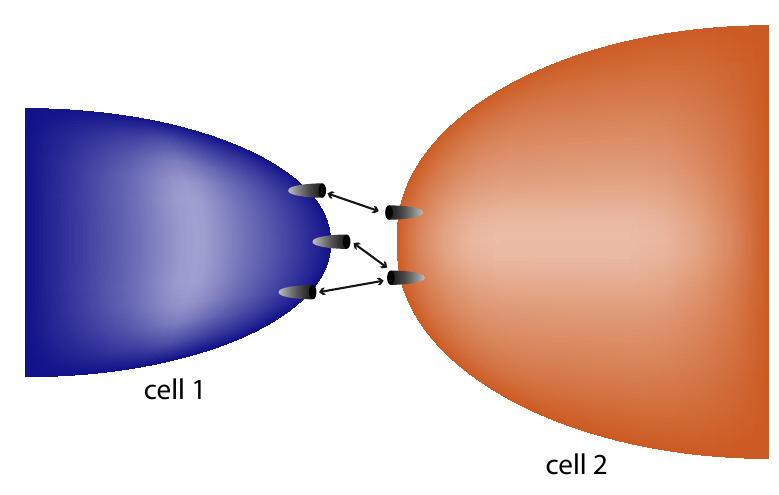}
\caption{
Discretized and smooth models of cellular morphology provide different resolutions of spatial relations between molecular complexes.
\newline
\textbf{Left:} Concentration-based simulation approaches that use discretized grid-like models of cellular morphologies are typically less computationally expensive than particle-based stochastic algorithms, but they cannot resolve length scales below the grid-constant. Consequently, the precise location of individual complexes on two apposing membrane-elements is unknown and only a very crude distance measure is available to determine if receptors are sufficiently close to each other to form bonds.
\textbf{Right:} In contrast, particle-based simulation on smooth grid-free cell geometry models can track the location of and distance between complexes and thus, bimolecular association possibilities based on spatial proximity arise naturally.
}\label{fig:discrete_vs_smooth_distance}
\end{figure}

\bigskip
The simulation platform Simmune is capable of creating computational models of juxtacrine signaling, provided that the biochemistry can be accurately captured by a deterministic dynamics \cite{angermann2012computational}. However, the assumptions that underly a deterministic description are not always justified \cite{rao2002control}. Additionally, in a deterministic  concentration-based context, only a very coarse measure of a gap distance is available and the above described interplay of the different length scales that influence the local biochemistry is hard to account for. This is because cell-cell contacts are computationally established by adjacent membrane and volume elements of the discretized cellular morphologies (Fig.~\ref{fig:discrete_vs_smooth_distance}, left). Because the reaction dynamics is concentration-based, each membrane complex residing on the same membrane element is assumed to be equally distant from each membrane complex located on the opposing membrane element, as the location of the complexes within one membrane element cannot be resolved. Similar remarks apply, in fact, to spatial lattice versions of the Stochastic Simulation Algorithm (SSA, frequently called \textit{Gillespie} algorithm) \cite{gillespie1976general, gillespie1977exact}. We emphasize that there are two fundamentally different spatial lattice based SSA methods: The lattice reaction-diffusion master equation (RDME) \cite{gardiner1976correlations} and the convergent reaction-diffusion master equation (CRDME) \cite{isaacson2013convergent, isaacson2018unstructured}. The CRDME overcomes certain limitations of the RDME approach, in particular, it converges to the continuous Doi-model \cite{doi1976second, doi1976stochastic, agbanusi2014comparison, prustel2017unified} in the limit of a vanishing grid-constant. Still, these spatial lattice methods are population-based as well and hence, as in the deterministic case, the location of individual complexes cannot be resolved up to the length scale set by the grid-constant. 

The situation is quite different, at least in principle, for particle-based computational models that track individual complexes. In such a setting, it is much easier to capture the various length scales and their potential regulatory impact, because interactions occur between individual molecules with well-defined locations (Fig.~\ref{fig:discrete_vs_smooth_distance}, right). 
Another, related advantage of particle-based methods is that they can resolve spatio-temporal correlations between a pair of just dissociated molecular complexes that give rise to a rebinding time distribution that can neither be accounted for by deterministic nor stochastic Gillespie approaches \cite{takahashi2010spatio}. It has further been shown that, in turn, these rebinding events can substantially modify the behavior of biochemical networks \cite{takahashi2010spatio}. While the CRDME, in principle, is capable of capturing such effects as well, to do so may require a highly resolved and hence computationally expensive spatial discretization.

To fully exploit these advantages of particle-based methods, several parts of the previously described algorithm had to be substantially modified as well. First, the applied computational modeling technique of cellular morphologies had to become flexible and sufficiently adaptable such that, for instance, even nanometer-sized contact sites established between T-cell microvilli and antigen-presenting cells (APCs) can be resolved. Second, the particle-based stochastic dynamics had to be able to include two-dimensional (2d) diffusion-influenced bimolecular reactions that are not only characterized by an encounter radius, but additionally an 'encounter height' to account for intermembrane binding reactions. Third, even free membrane diffusion had to be altered by assigning a 'steric height' to molecular components of membrane complexes that prevent them from diffusing into the specialized sites of contact between two cells.   

Other software tools, such as \cite{tolle2010meredys, gruenert2010rule, fange2010stochastic, sneddon2011efficient, hepburn2012steps, arjunan2013cell, roberts2013lattice, drawert2016stochastic, michalski2016springsalad, blinov2017compartmental, andrews2017smoldyn, isaacson2018unstructured, hoffmann2019readdy, sokolowski2019egfrd, varga2020nerdss, husar2022mcell4} provide either spatially-resolved or particle-based stochastic simulation capabilities or both, but, to our knowledge, the approach presented here is the first to combine those aspects for general geometries in a grid-free manner. 
We point out that Smoldyn \cite{andrews2017smoldyn} and MCell4 \cite{husar2022mcell4}  have been reported to support intermembrane reactions as well even though there have – to our knowledge – not been any applications of those capabilities.
Ref.~\cite{johnson2021quantifying} discusses in more detail deterministic and stochastic non-spatial and spatially-resolved simulation approaches and software tools, while the review \cite{andrews2018particle} focuses on particle-based stochastic simulators.    

In the following, we will describe the theoretical basis of the new algorithm. The remainder of this manuscript is organized as follows. 
In Sec.~\ref{comp_model_geo} we describe an extension of the previously utilized well-known implicit surface modeling technique, referred to as metaballs, that increases its adaptability.
Sec.~\ref{stochastic_dynamics} briefly introduces particle-based stochastic simulation algorithms that are based on the Smoluchowski-Collins-Kimball (SCK) \cite{rice1985diffusion} model of diffusion-influenced reactions. Then, we discuss the integration of these algorithms with the computational representation of model geometries and, in particular, the necessary extensions to accommodate bimolecular reactions that involve complexes from two adjacent membranes.
Finally, in Sec.~\ref{applications} we present applications of the proposed algorithm. 
\begin{figure}
\includegraphics[scale=0.75]{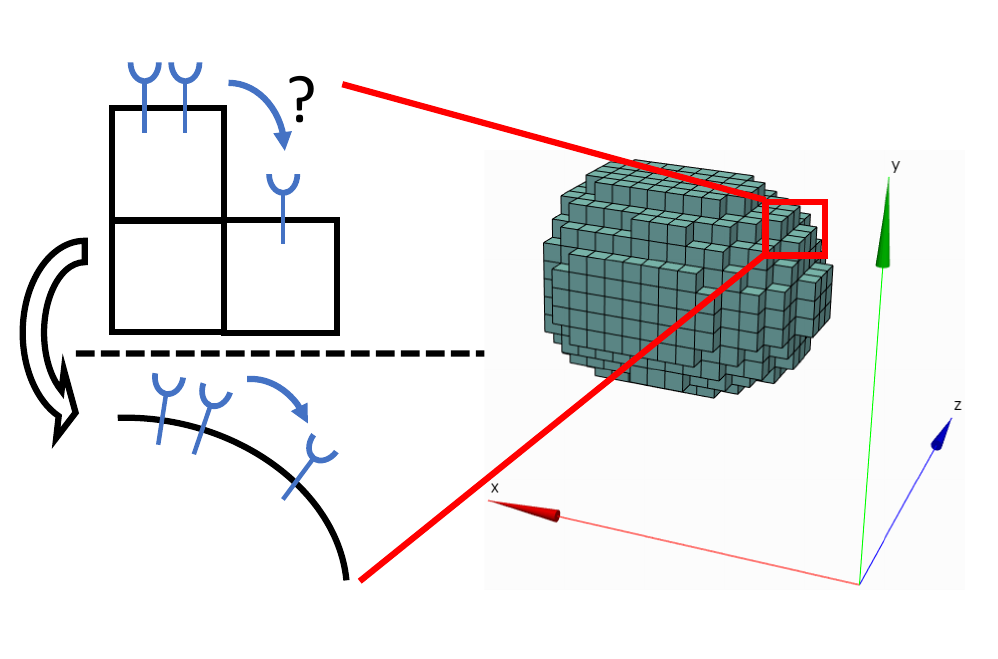}
\caption{
Schematic illustration of the technical challenges associated with combining discretized cube-based models of cellular morphology and trajectory-based stochastic simulations of molecular diffusion. These challenges are particularly severe for diffusion on a surface. Here, the time-stepping algorithm that generates the molecules' position updates relies critically on geometric constructs, such as the local tangent space, that can readily be obtained for a smooth surface. By contrast, these required geometric objects are classically not well-defined at the edges and corners that come with the stair-like structure of discretized surfaces.
}\label{fig:diffusion_over_edge}
\end{figure}
\section{Computational models of cell-cell contact site geometries}  
\label{comp_model_geo}
Many deterministic discretizations of reaction-diffusion processes employ finite volume techniques with, frequently cubic, volume elements, called voxels. Using deterministic models for the time-evolution of the biochemistry, straightforward numerical corrections can be employed to control cubic discretization artifacts \cite{novak2007diffusion, angermann2012computational}. However, it is much more difficult to reconcile a discretization of space with stochastic particle-based dynamics than with a concentration-based deterministic approaches.
The stochastic formulation specifies the system's state by tracking all particle locations. The equations of motion are stochastic differential equations \cite{van1992stochastic, gardiner2009stochastic} that govern the particles' position increments and thus generate the continuous trajectories (Brownian paths) that describe the diffusive motion of individual molecular complexes in time and space.
To coexist with a cubic discretization of space, the stochastic dynamics would need to be profoundly modified.
 For the case of surface diffusion, this effort would involve finding a generalization of the Brownian position updates such that particles' paths even around corners and edges of voxels become well-defined (Fig.~\ref{fig:diffusion_over_edge}). The reason why this is challenging is that stochastic dynamics on curved surfaces invokes differential geometric constructs, such as the local tangent space, to be able to specify positions updates, as we will see in more detail in Sec.~\ref{stochastic_dynamics}. However, these mathematical objects are not well-defined at edges and corners in classical differential geometry. Consequently, as a prerequisite for implementing stochastic dynamics on discretized surfaces, one first would need to identify and apply a suitable discrete differential geometric formulation.

To avoid these technical challenges, we previously employed \cite{prustel2020stochastic} metaballs \cite{blinn1982generalization, gomes2009implicit, de2015survey, akenine2018real}, also referred to as blobs, a well-known implicit surface modeling technique that provides realistic smooth and grid-free computational representations of spatial domains.
\begin{figure}
\includegraphics[scale=0.5]{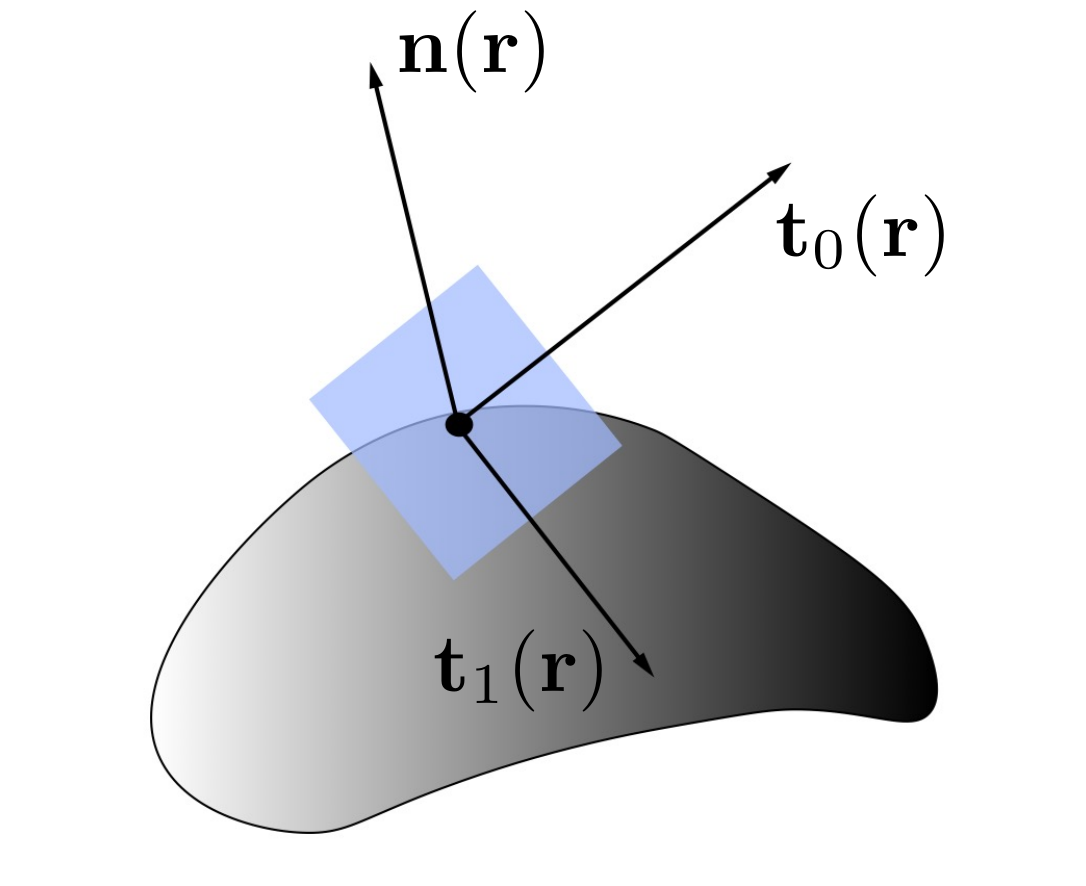}
\includegraphics[scale=0.3]{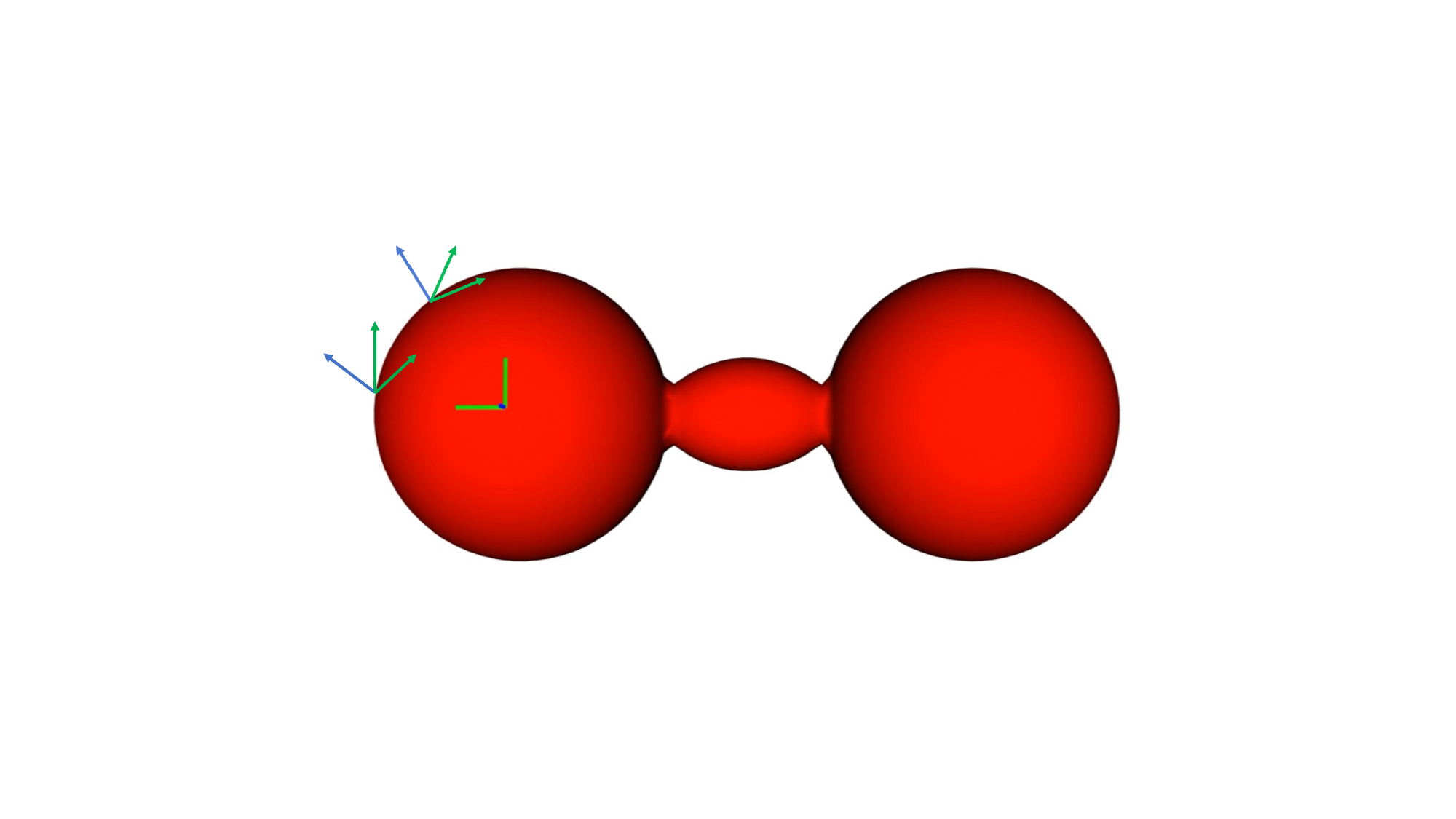}
\caption{
The local tangent space of a smooth surface.
\newline
\textbf{Top:}
Schematic depiction of the tangent vectors $\mathbf{t}_{0}(\mathbf{r})$, $\mathbf{t}_{1}(\mathbf{r})$ and the normal $\mathbf{n}(\mathbf{r})$ that furnish a basis of the local tangent space at a point $\mathbf{r}$ of a grid-free represented surface.
\textbf{Bottom:} Snapshot of three Brownian walkers that represent diffusing molecular complexes on a metaball-generated surface. The shown local tangent spaces at the walkers' current locations are a prerequisite for specifying the walkers' next position updates (Eq.~(\ref{free_membrane_propagation})). Here, the blue and green arrows depict the normal and tangent vectors, respectively. Geometric models composed of metaballs offer the advantage that the local tangent space can efficiently be computed.
}\label{fig:local_frame}
\end{figure}
Metaballs are specified by a potential, that is, a scalar function in three-dimensional (3d) space
\begin{eqnarray}
\varphi &:& \mathbb{R}^{3} \rightarrow \mathbb{R}, \nonumber\\
&& \mathbf{r}=(x, y, z) \rightarrow \varphi(\mathbf{r}),
\end{eqnarray}
More concretely, here we use a potential that assumes the form
\begin{equation}\label{potential}
\varphi(\mathbf{r}) = \left\{ 
\begin{array}{ll}
 & \left[1-\left(\frac{\vert \mathbf{r} - \mathbf{r}_{0}\vert}{R}^2\right) \right]^{2} \qquad \mathrm{if}\, \vert \mathbf{r} - \mathbf{r}_{0}\vert < R\\ & 0 \qquad \qquad \qquad \quad \mathrm{otherwise}, 
 \end{array} \right.
\end{equation}
where $R$ and $\mathbf{r}_{0}$ refer to the metaball's radius and position, respectively. Also, we use the notation $\vert\mathbf{r}\vert=\sqrt{\mathbf{r}\cdot\mathbf{r}}$ for the length of a vector $\mathbf{r}$. Note that this potential's influence is local as it vanishes outside the finite domain given by $\vert \mathbf{r} - \mathbf{r}_{0}\vert < R$.

Metaballs entail the major advantage that more complex geometries can readily be created by superposing their potential functions, a technique that is known as blobby modeling \cite{blinn1982generalization}.
The resulting object is then described by the function
\begin{equation}
\label{blobby_blend}
\phi(\mathbf{r}) = s - \sum_{\text{metaballs}}\varphi_{i}(\mathbf{r}), 
\end{equation}
where $s > 0$ is a threshold value that can be freely chosen.
\begin{figure}
\begin{subfigure}[t]{0.575\textwidth}
\includegraphics[scale=0.2]{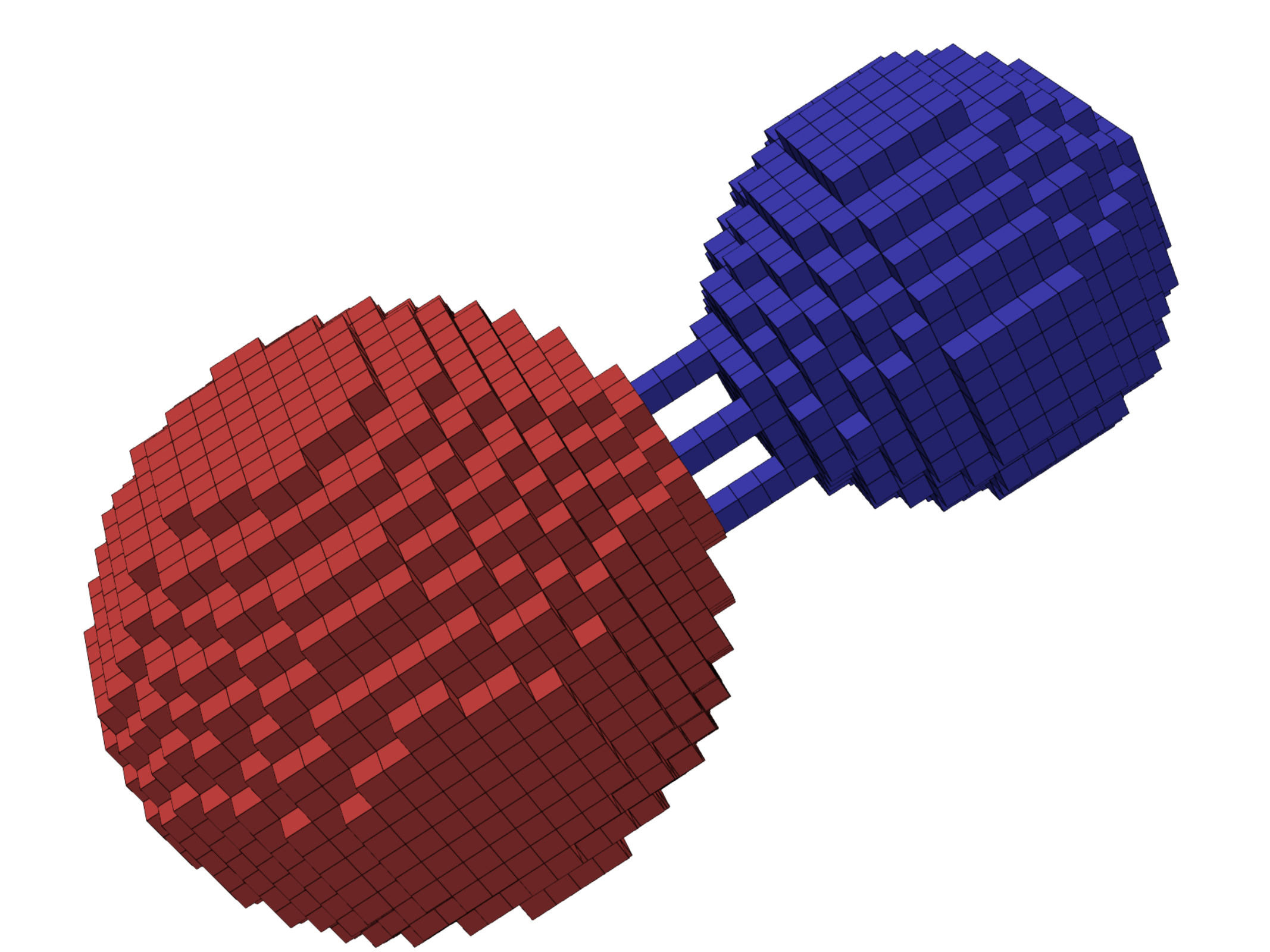}
\caption{
Example of a grid-based finite volume representation of two adjacent cells forming several cell-cell contact sites. Such discretized models are typically used in concert with concentration-based deterministic descriptions of biochemistry.
}
\label{sub_fig:voxel_rep}
\end{subfigure}
~
\begin{subfigure}[t]{0.575\textwidth}
\includegraphics[scale=0.2]{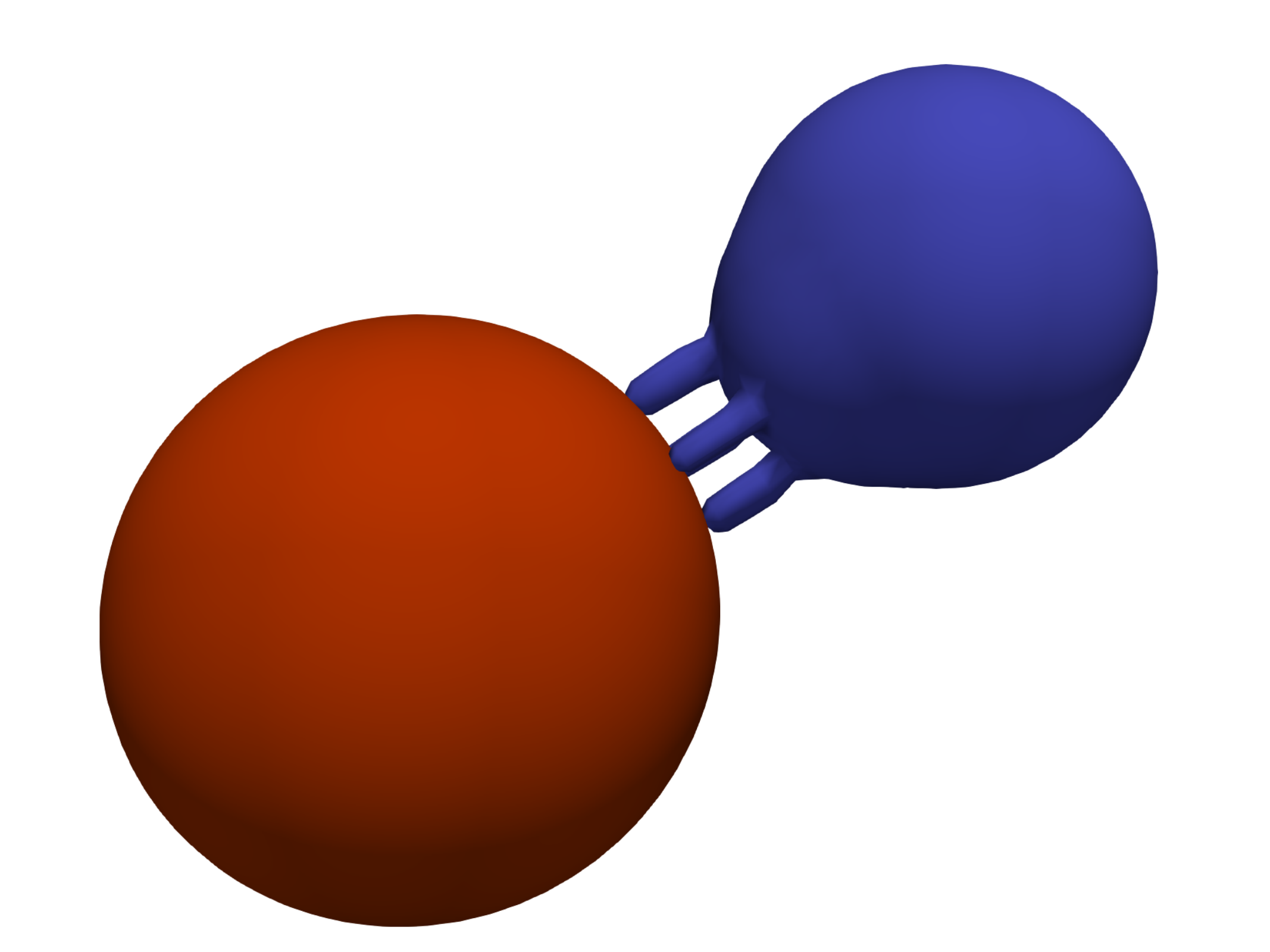}
\caption{
The corresponding smooth and grid-free metaball-based model is suitable for particle-based stochastic approaches to cellular biochemistry. At the same time, it can be used as an auxiliary data structure to generate the discretized representation displayed above.
}
\label{sub_fig:smooth_rep}
\end{subfigure}
\caption{
Discrete and corresponding smooth models of cellular morphology.
}\label{fig:voxel_vs_blob_rep}
\end{figure}

\begin{figure}
\includegraphics[scale=0.175]{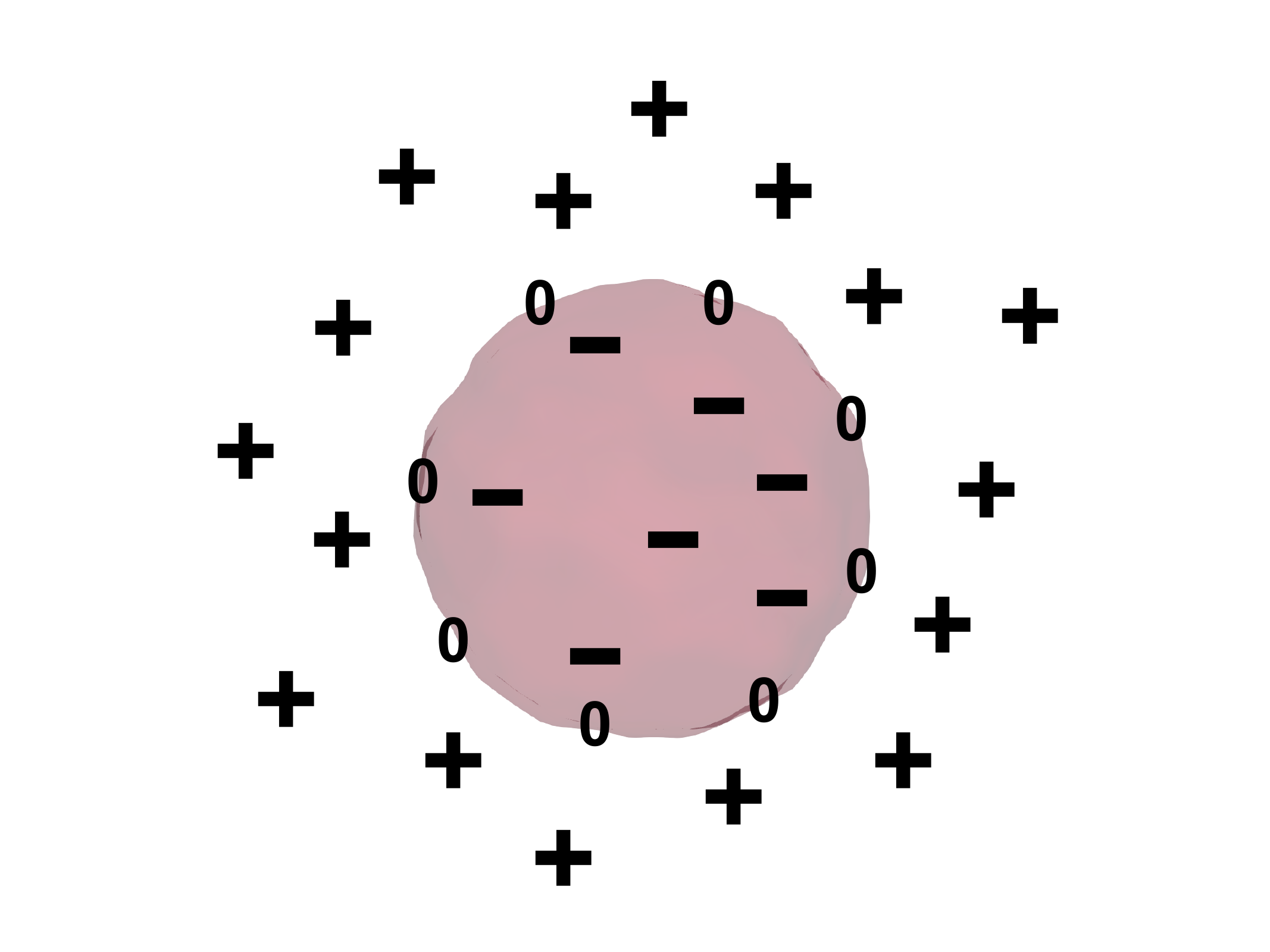}
\includegraphics[scale=0.175]{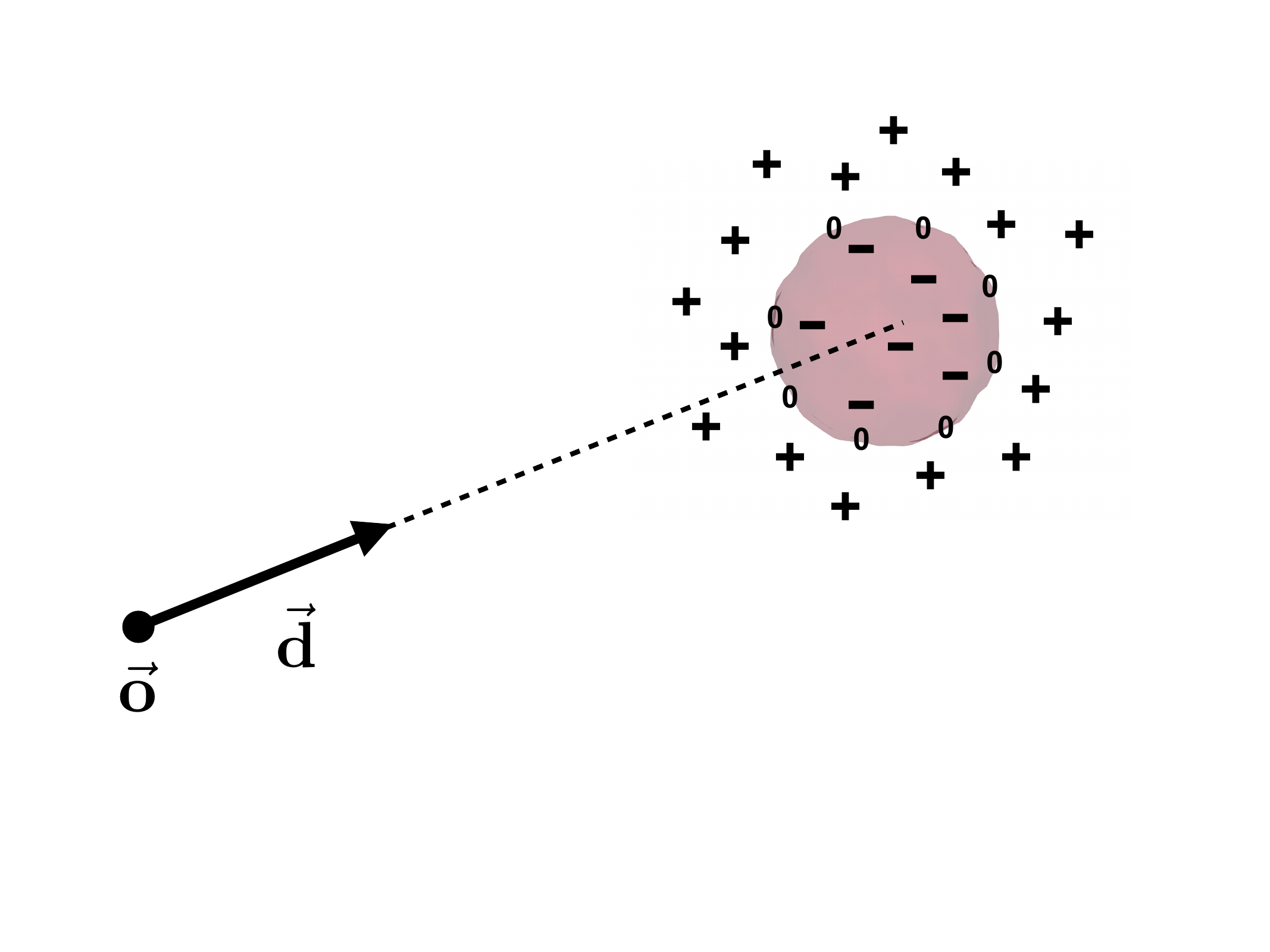}
\caption{
Features of metaball-based modeling that facilitate the identification of those space regions that are occupied by cells.
\newline
\textbf{Left:} Metaballs enable swift checks whether a point $\mathbf{r}$ lies inside ($\phi(\mathbf{r}) < 0$), outside ($\phi(\mathbf{r}) > 0$) or on the surface ($\phi(\mathbf{r}) = 0$) of the represented object. \textbf{Right:} Ray-object intersection is detected by  a sign change (Eq.~\ref{sign_check}) along the ray. The intersection point on the surface $\phi(\mathbf{r}_{\text{ray}}(\lambda)) = 0$ can be determined by bisection.
}\label{fig:metaball_in_out}
\end{figure}

\begin{figure}
\begin{subfigure}[t]{0.575\textwidth}
\includegraphics[scale=0.2]{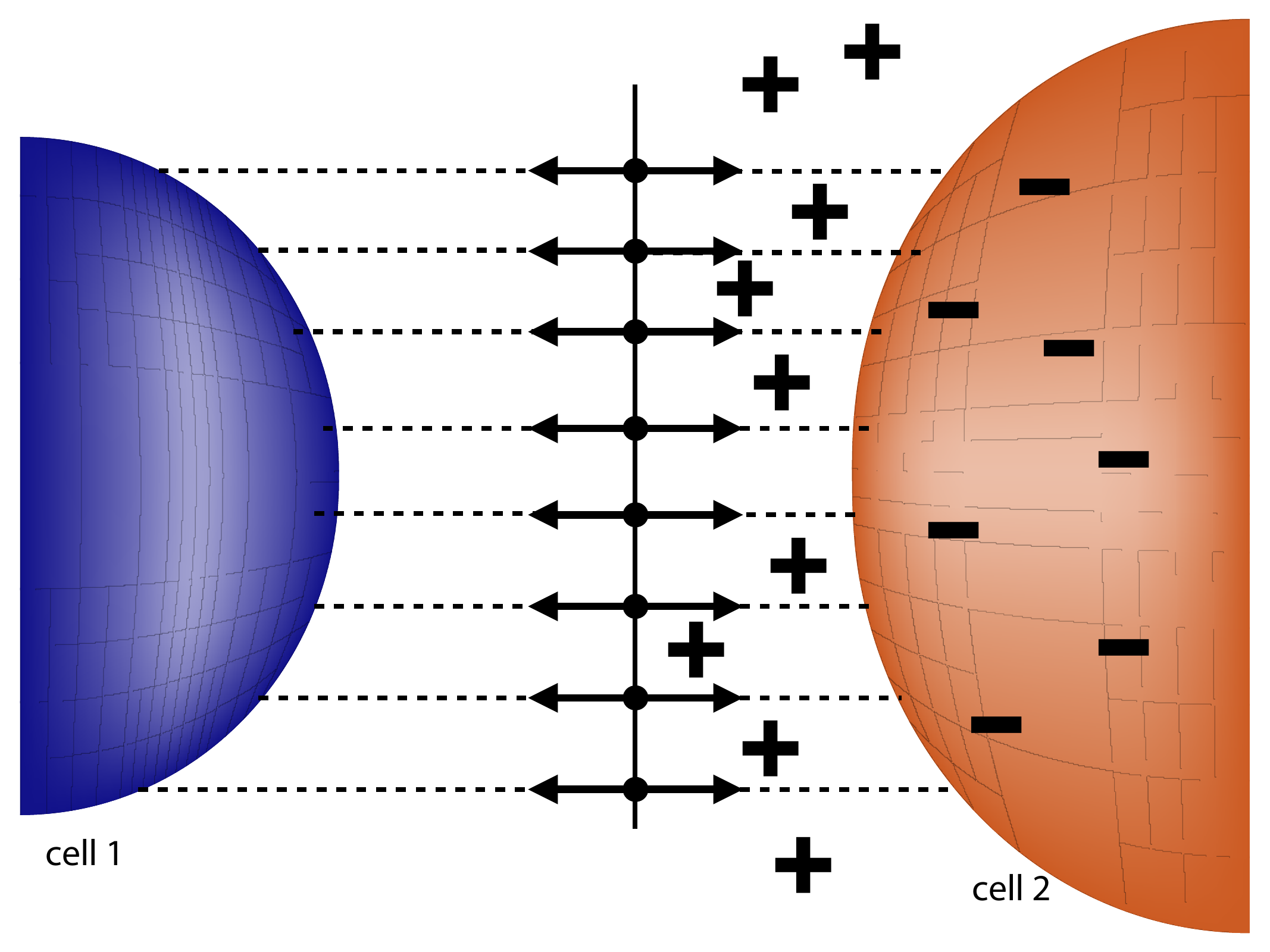}
\caption{
A set of rays can be used to measure the distances between points of the two membrane regions that form the contact site. 
}
\label{sub_fig:gap_distance}
\end{subfigure}
~
\begin{subfigure}[t]{0.45\textwidth}
\includegraphics[scale=0.285]{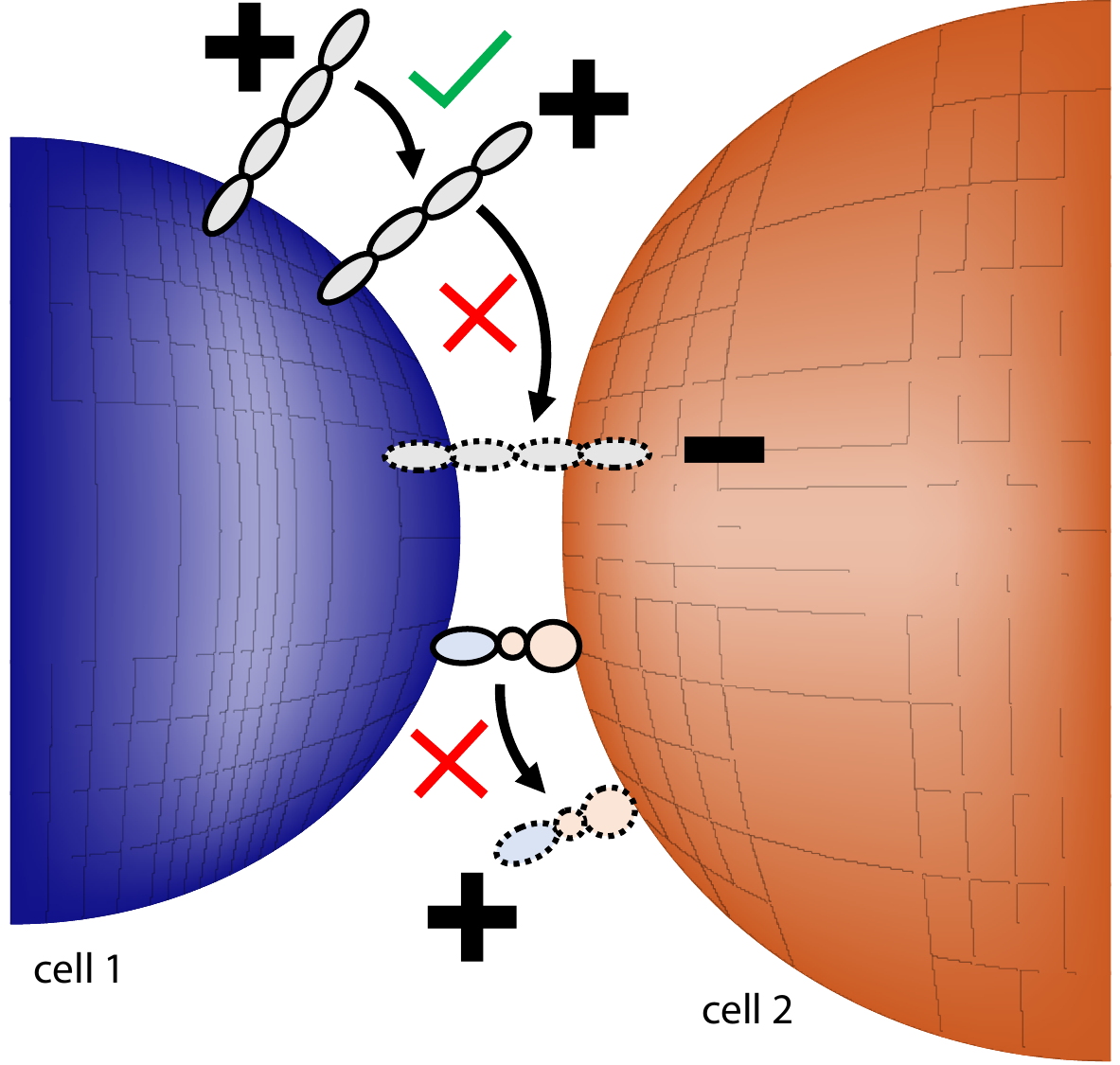}
\caption{
Constraints imposed by the contact site's morphology on diffusion of complexes with a steric height and of intermembrane complexes that span the gap can be efficiently implemented by application of Eq.~(\ref{sign_check}). 
}
\label{sub_fig:constrained_prop}
\end{subfigure}
\caption{
Instances of intercellular distance measurements and checks that are used by the algorithm and that rely on blobby modeling techniques.  
}\label{fig:blobby_check_apps}
\end{figure}

Another crucial advantage offered by blobby modeling is that it provides simple and efficient criteria to determine whether a point $\mathbf{r}$ is located inside or outside of the represented object. Similarly, the object's surface is given by the 0-isocontour (Fig.~\ref{fig:metaball_in_out}). More concisely, this reads as 
\begin{equation}
\label{sign_check}
\phi(\mathbf{r}) = \left\{ 
\begin{array}{lll}
 & > 0  \qquad \text{if} \,\, \mathbf{r} \,\, \text{is outside object}\\ 
 & < 0 \qquad  \text{if} \,\, \mathbf{r} \,\, \text{is inside object} \\
 & = 0  \qquad \text{if} \,\, \mathbf{r} \,\, \text{is on object's surface}.
 \end{array} 
 \right.
\end{equation}
This ability of blobby modeling plays an essential role in many parts of our simulation algorithm, such as particle propagation, distance measurements between cells (Fig.~\ref{sub_fig:gap_distance}) and imposing geometric constraints (Fig.~\ref{sub_fig:constrained_prop}). In contrast, modeling techniques based on explicit parameterizations do not offer such an efficient check.

A grid-free metaball-based representation (Fig.~\ref{sub_fig:smooth_rep}) is automatically generated by Simmune \cite{angermann2012computational} as an purely auxiliary device to help create its discretized counterpart (Fig.~\ref{sub_fig:voxel_rep}). In the context of stochastic simulations, this auxiliary data structure is promoted to the actual computational representation of cellular geometry and can directly be used to create stochastic model soluble ligand mediated cellular signaling processes. In this way, both a discretized grid and smooth grid-free geometry representation
well-suited for deterministic and stochastic simulations, respectively, can be obtained without any extra effort. Moreover, the close
relationship between these two models of geometry allows us to compare deterministic and stochastic simulation results.

However, the case of cell-cell contact interfaces typically requires further adjustments.
While its discrete counterpart can only describe whether two membrane voxels of two different cells are at contact with each other, the smooth model provides a substantially higher resolved description of the cell-cell interface. The intermembrane gap distance is an instance of such a higher resolved quantity that becomes relevant in the context of a smooth geometry and that, therefore, one should be able to accurately adjust to reflect biological reality. To impose a chosen gap distance on a single contact site, translating one of the two cells in its entirety is sufficient. However, this simple strategy fails for several contact sites (Fig.~\ref{sub_fig:smooth_rep}), because one translation typically cannot satisfy multiple gap distance requirements simultaneously. 

\begin{figure}
\includegraphics[scale=0.25]{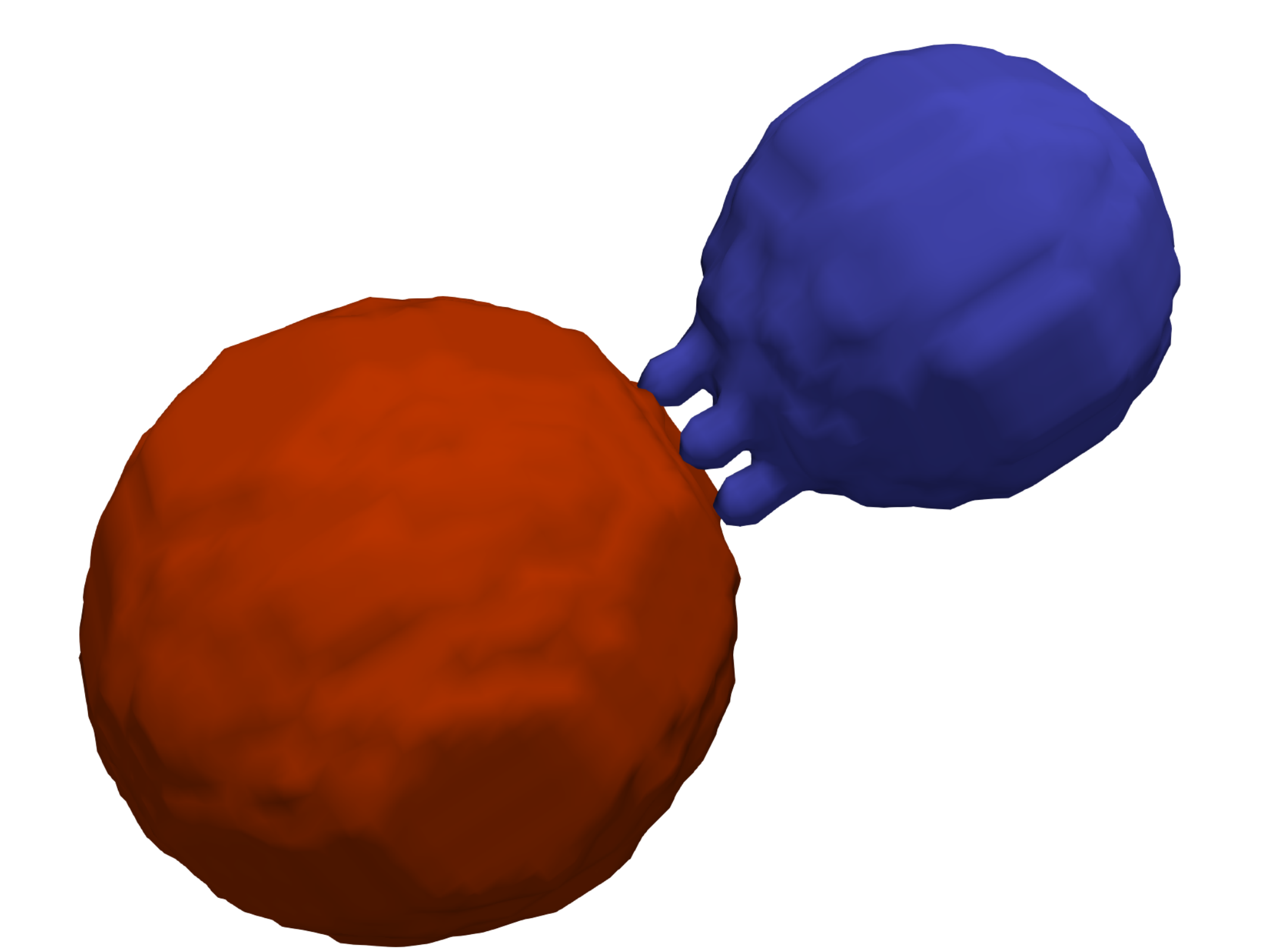}
\caption{
An alternative smooth metaball-based representation is shown that is distinguished from the previous one (Fig.~\ref{sub_fig:smooth_rep}) by a higher degree of adaptability. To construct this model, the smooth model shown in Fig.~\ref{sub_fig:smooth_rep} once again serves as the starting point. However, the voxels that would generate the discretized version are now replaced by metaballs. Then, the totality of all such created metaballs gives rise to the cellular model depicted here.  
}\label{fig:alternative_smooth_rep}
\end{figure}

To address this issue, we perform a transformation that, starting from the discretized representation, replaces each of its voxels by a metaball given by Eq.~(\ref{potential}) located at the voxel's center. This procedure again results in a grid-free representation of the cell (Fig.~\ref{fig:alternative_smooth_rep}).
The major advantage is the model's increased flexibility and adaptability one has thereby acquired, as the position and radius of all the individual metaballs can be independently adjusted. Now, the alternative representation can satisfy locally different distance requirements at multiple contact sites. Importantly, this new metaball-based model is still closely related to its discretized counterpart, thereby preserving the ability to compare deterministic and stochastic simulations readily. 

However, the added adaptability comes at a price in principle. Because the number of metaballs typically increases substantially, the brute-force calculation of the sum of all their potentials (Eq.~(\ref{blobby_blend})) becomes more expensive. Fortunately, as an immediate remedy, one may only consider the influence of metaballs in a neighborhood of a given point of interest. This is made possible by the already mentioned locality and cut-off property of the potential function that we use (Eq.~(\ref{potential})) to specify the metaballs.

Now, to quantitatively adjust the intermembrane gap, we use rays, semi-infinite lines, given by
\begin{equation}
\label{ray_def}
\mathbf{r}_{\text{ray}}(\lambda) = \mathbf{o} + \lambda \mathbf{d},
\end{equation}
where $\mathbf{o}$ and $\mathbf{d}$ are the ray's origin and its (unit) direction vector, respectively. Rays are widely used in rendering algorithms \cite{akenine2018real}. Here, we use an ensemble of rays and a simple ray-object intersection algorithm (Fig.~\ref{fig:metaball_in_out}) to collect a set of distances between the two contact sites (Fig.~\ref{sub_fig:gap_distance}). The minimum of these distances represents the intermembrane gap distance. In this way, we can also guide the adjustments to the metaballs, that form the contact site, to change the intermembrane gap distance to a desired value.  
The ray-object intersection algorithm provides another example of the benefits of the metaball representation. Without the efficient way of determining whether a point lies inside or outside of a cell, the intersection and subsequent bisection would become much more expensive. 

\begin{figure}
\begin{subfigure}[t]{0.56\textwidth}
\includegraphics[scale=0.2]{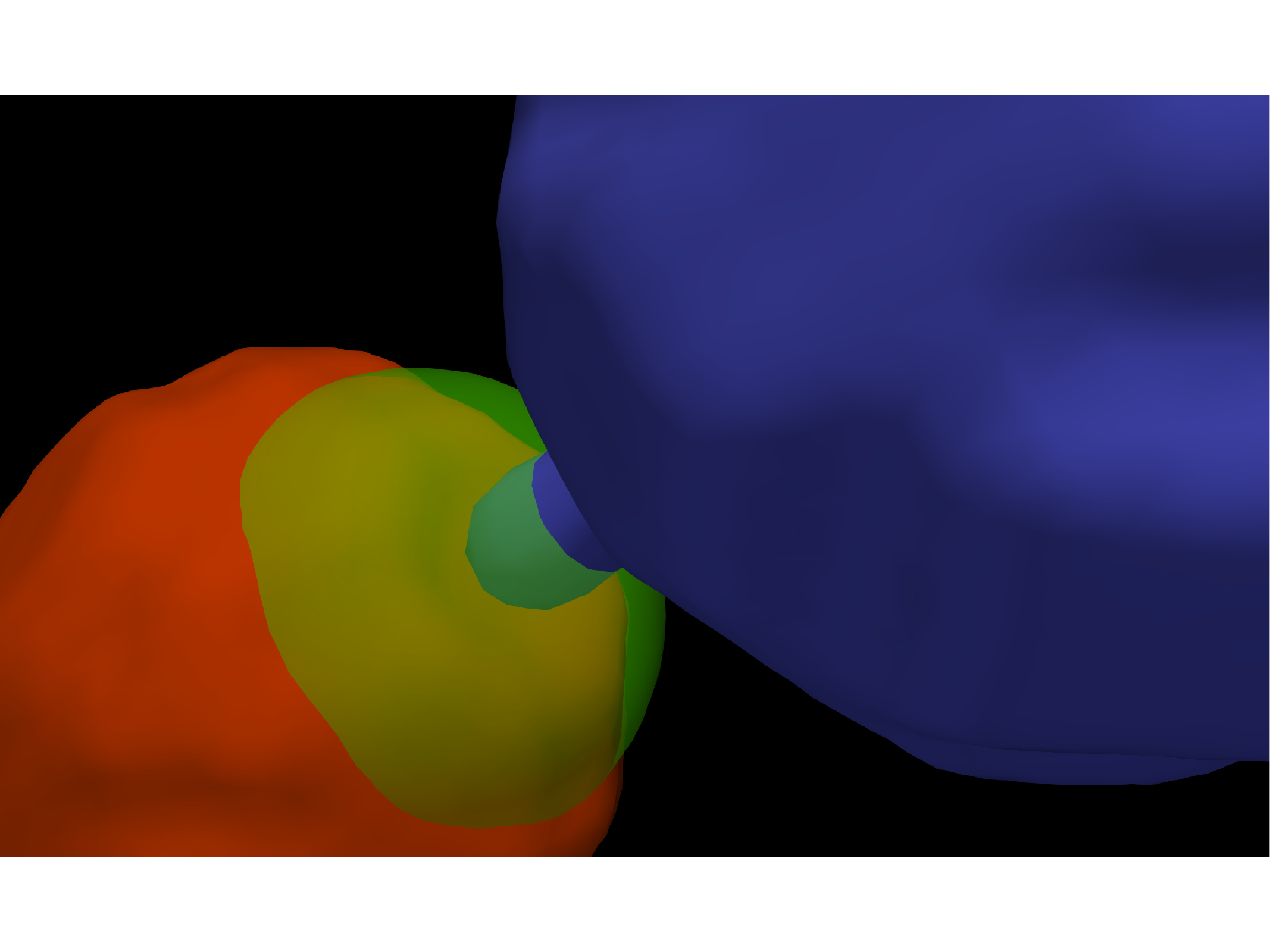}
\caption{
The green-colored domain indicates a metaball that does not represent a cell, but that instead defines a portion of the cell's membrane by intersecting the blobs that comprise the cell.
}
\label{sub_fig:init_distribution_region}
\end{subfigure}
~
\begin{subfigure}[t]{0.56\textwidth}
\includegraphics[scale=0.2]{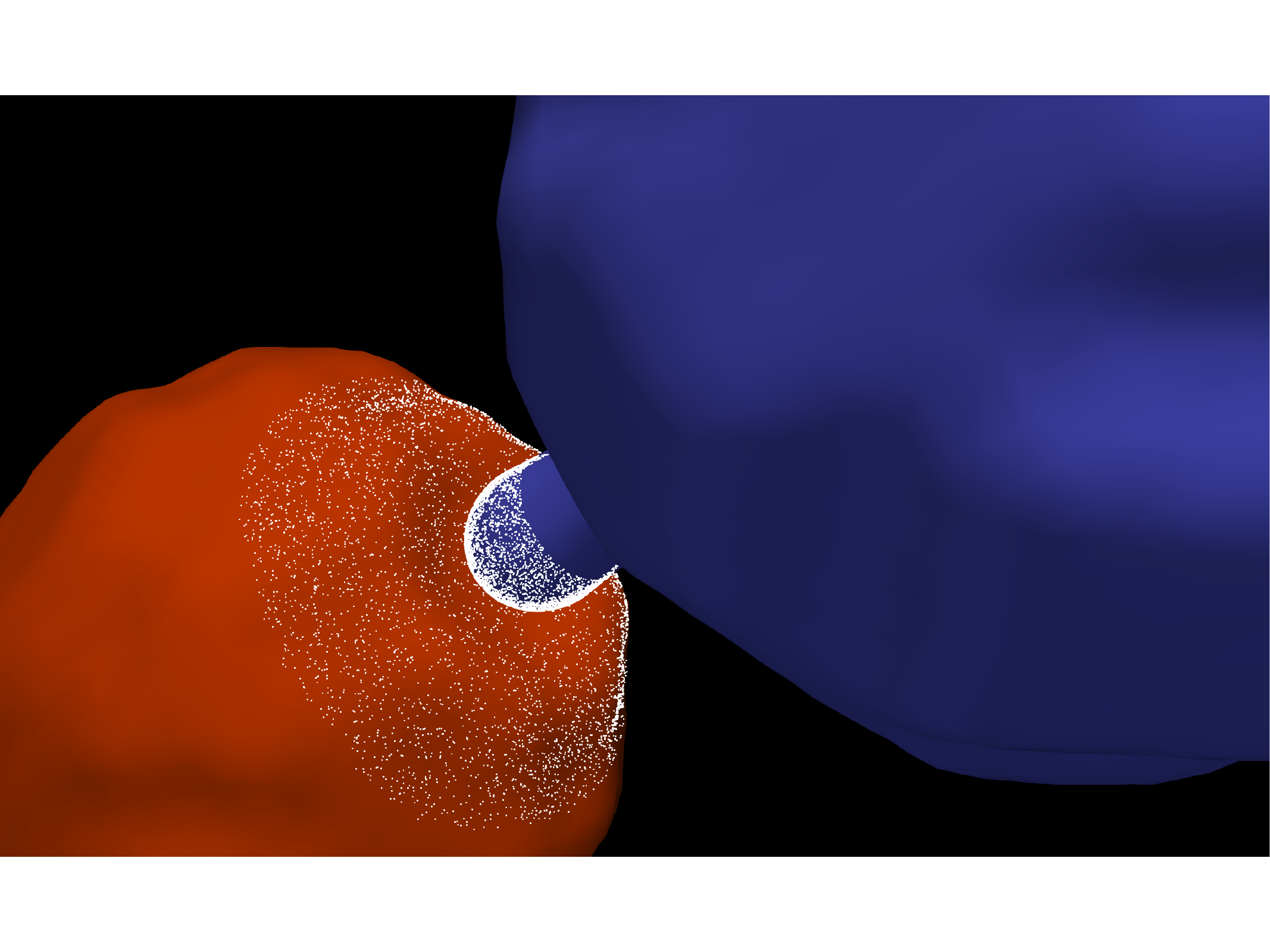}
\caption{
The white dots represent the initial positions of the molecular complexes that now populate the previously specified membrane region (Fig.~\ref{sub_fig:init_distribution_region}).
}
\label{sub_fig:init_position_distribution}
\end{subfigure}
\caption{
Specification of the initial spatial distribution of membrane complexes.
}\label{fig:init_distribution}
\end{figure}

Finally, we point out that intersections of metaballs can easily be described, again by exploiting Eq.~(\ref{sign_check}). Thus, the cell regions (Fig.~\ref{sub_fig:init_distribution_region}) that are initially populated by complexes (Fig.~\ref{sub_fig:init_position_distribution}) can be specified. 
\section{Stochastic modeling of biochemistry at cell-cell contacts}
\label{stochastic_dynamics}
\begin{figure}
\begin{subfigure}[t]{1.\textwidth}
\includegraphics[scale=0.25]{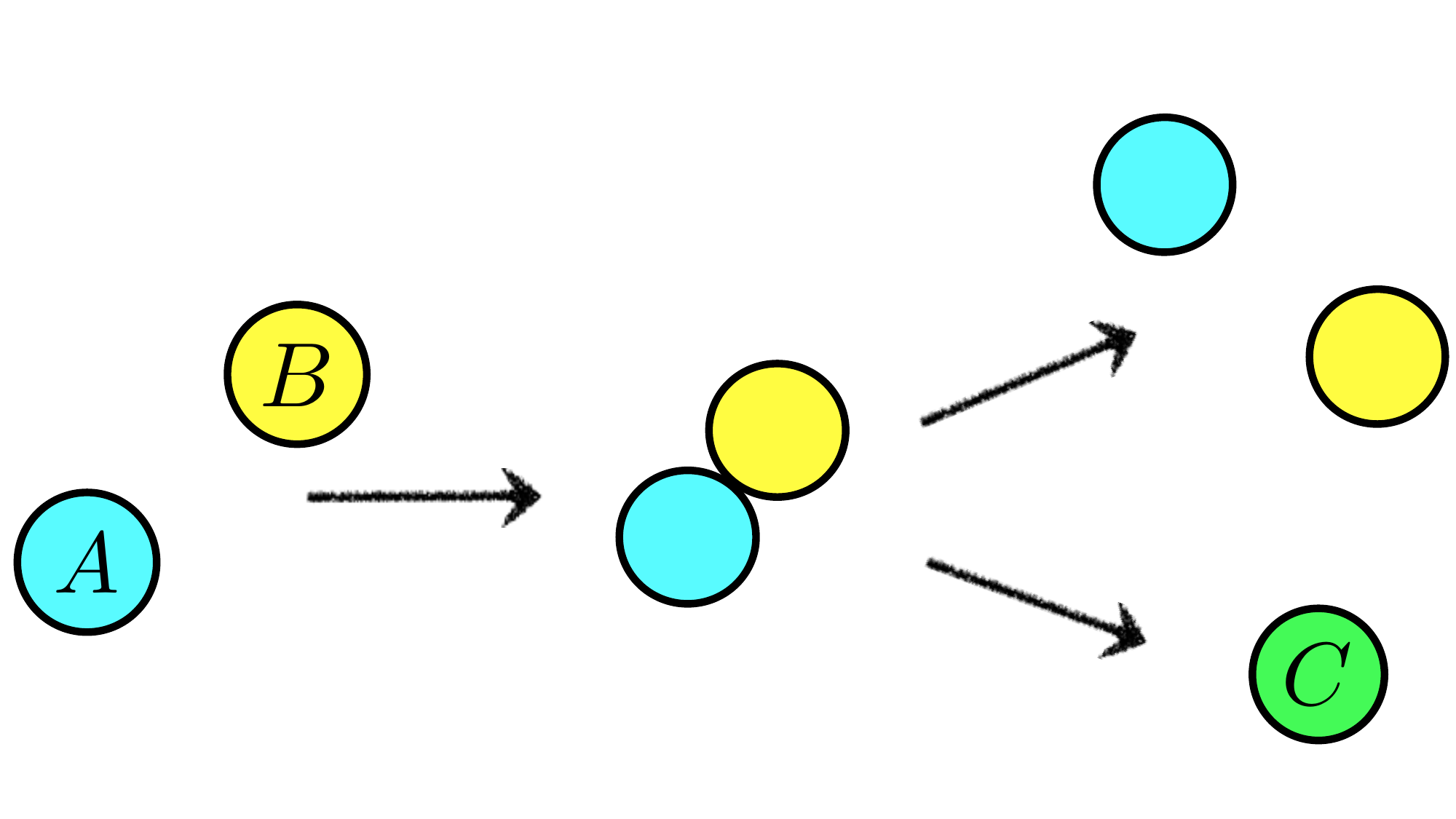}
\caption{
The Smoluchowski picture describes diffusion-influenced bimolecular reactions $A+B\rightarrow C$ as a two-step process, where a pair of molecules first encounter each other due to their diffusive motion, resulting either, after possibly many subsequent encounters, in a bound state or escape from each other without binding. 
}
\label{sub_fig:sck_two_step}
\end{subfigure}
~
\begin{subfigure}[t]{1.\textwidth}
\includegraphics[scale=0.25]{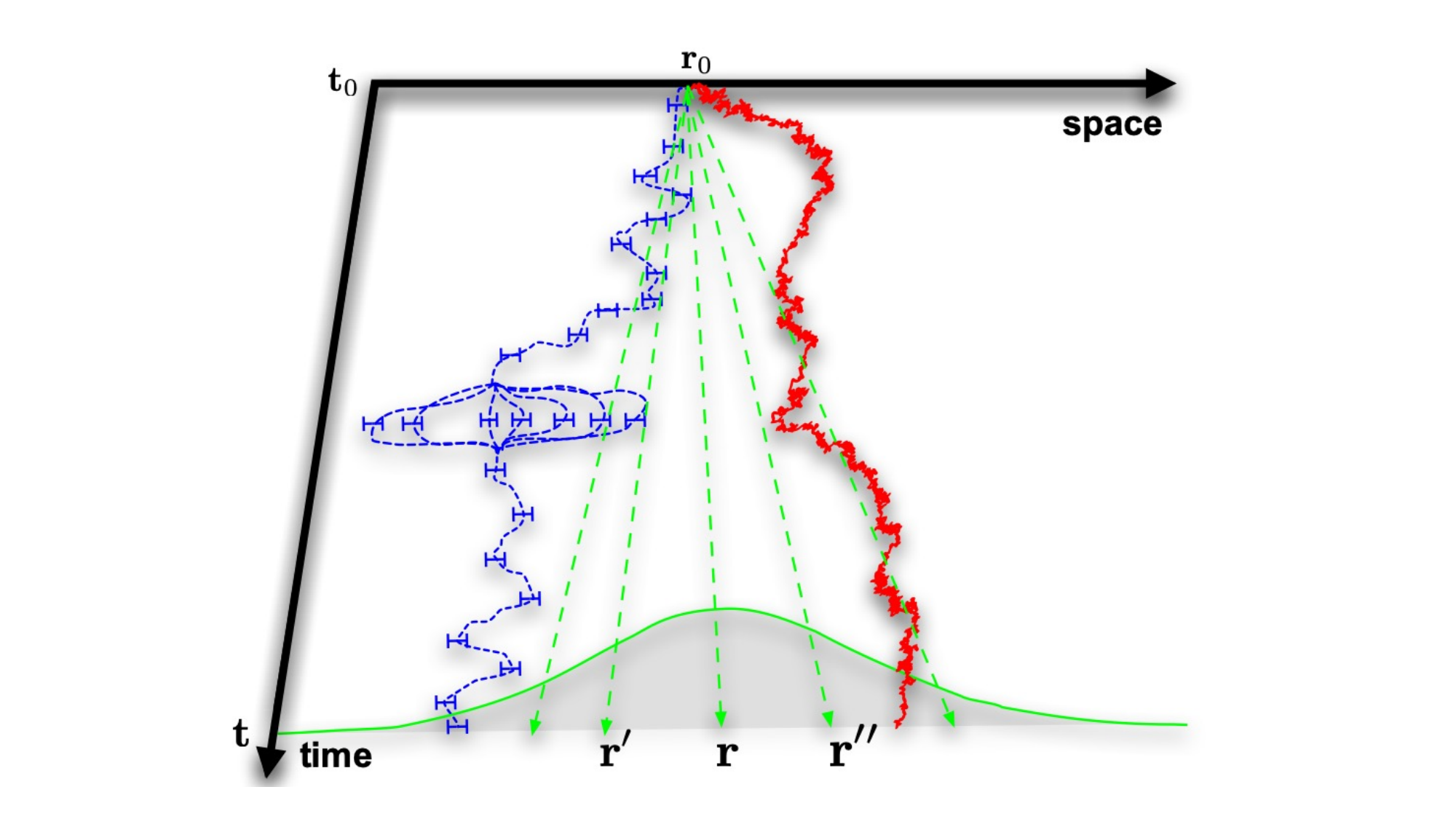}
\caption{
Diffusive motion of molecules can be described by different yet interrelated and equivalent mathematical formulations. A solution of the Langevin equation is a trajectory-valued random variable. Simulations of the Langevin equation produce stochastically generated paths as specific realizations of this random variable. They amount to sampling from the probability functional on the space of all possible trajectories between initial and terminal points $\mathbf{r}_{0}$, $\mathbf{r}$ passing through domains at certain times (blue path). However, a numerical realization obtained from one simulation provides a time-discretized version of the true Brownian trajectory (red path), specifying the position at certain times only. The need to minimize that uncertainty lies at the root of the notorious inefficiency of Brownian Dynamics simulations when employed to study diffusion-influenced reactions and motivate the use of Green's functions, distinct solutions of the diffusion equation, as a remedy. Green's functions represent the conditional probability to find the molecule at $\mathbf{r}$ at time $t$, given that initially it was located at $\mathbf{r}_{0}$ (the grey region targeted by the green arrows).
}
\label{sub_fig:faces_diffusion}
\end{subfigure}
\caption{
Theoretical approaches to diffusion and diffusion-influenced bimolecular reactions.
}\label{fig:diffusion_foundations}
\end{figure}

We consider particle-based stochastic simulation algorithms that are based on the SCK model \cite{smoluchowski1917mathematical, gosele1984reaction, rice1985diffusion} of diffusion-influenced bimolecular reactions (Fig.~\ref{sub_fig:sck_two_step}). While in Sec.~\ref{applications} we will use specifically the algorithm described in Ref.~\cite{prustel2021space}, the SCK framework is widely used in a number of stochastic simulation algorithms to capture the diffusive component of bimolecular reactions \cite{edelstein1993brownian, edelstein1997brownian, kim1999dynamic, barenbrug2002accurate, van2005simulating, oppelstrup2009first, johnson2014free, sokolowski2019egfrd}.

Different, yet equivalent, theoretical approaches (Fig.~\ref{sub_fig:faces_diffusion}) have been used to implement the diffusion component of the SCK model in simulation algorithms. The Langevin stochastic differential equation \cite{van1992stochastic, gardiner2009stochastic} furnishes the theoretical basis of Brownian Dynamics algorithms and accounts for the time evolution of the molecules' positions directly in terms of particle trajectories. Solutions of the Einstein-Smoluchowski diffusion equation \cite{van1992stochastic, gardiner2009stochastic}, by contrast, are conditional probability density functions $p(\mathbf{r}, t \vert \mathbf{r}_{0})$ that yield the probability of finding a molecule at a position $\mathbf{r}$ at time $t$, provided it was sharply localized at  $\mathbf{r}_{0}$ at time $t=0$. The probability density functions that are conditioned on an initially sharply localized particle are also referred to as Green's functions or propagators.  

To overcome the high cost of simulations based solely on Brownian Dynamics simulations, many of the aforementioned algorithms adopt a hybrid approach and invoke Green's functions to enhance trajectory-based simulations. Because of these commonalities, most of the adaptations necessary to integrate models of cellular geometry and stochastic dynamics that we will discuss in the following apply generally to the various stochastic algorithms.

Particle-based simulation approaches have originally been developed for flat Euclidean spaces and hence tacitly rely on features specific to these geometries, such as the existence of a global coordinate system. Hence, at least those parts of the algorithms that involve spatial aspects, in particular particle propagation, need to be generalized to become applicable to more general geometries. The most profound adaptions are necessary for the diffusion and interactions between membrane complexes.
This is in particular true in the presence of cell-cell contact sites where novel features emerge that can impact the local biochemistry and even free membrane diffusion. Therefore, we start by recalling how free diffusion can be implemented on metaball-based curved surfaces \cite{prustel2020stochastic}.
\subsection{Free membrane diffusion}
The essential strategy is to draw on the locally Euclidean structure of the local tangent spaces, also referred to as frames, that can be defined at any point $\mathbf{r}_{0}$ of a sufficiently well-behaved surface (Fig.~\ref{fig:local_frame}). Thus, equations of motion devised for globally Euclidean spaces, can immediately be defined locally in a sufficiently small neighborhood of $\mathbf{r}_{0}$. The remaining challenge is to piece together the dynamic updates in the local patches consistently. This task has been accomplished \cite{holyst1999diffusion} for free diffusion on a surface that can be characterized by
\begin{equation}\label{implicit_surface}
\psi(\mathbf{r}) = 0,
\end{equation}
where $\psi$ denotes a scalar function on $\mathbb{R}^{3}$ or a subset thereof. Then, the local tangent space at a point $\mathbf{r}_{0}=(x_{0}, y_{0}, z_{0})$ is the solution to
\begin{equation}
(\mathbf{r} - \mathbf{r}_{0}) \cdot \nabla\psi(\mathbf{r}_{0}) = 0,
\end{equation}
where the gradient $\nabla\psi(\mathbf{r}_{0})$ is the surface's normal at point $\mathbf{r}_{0}$. Having the tangent space at one's disposal, one can define a free Brownian position update reminiscently to the globally Euclidean case as
\begin{equation}\label{free_membrane_propagation}
\mathbf{r}^{\ast}(t+\Delta t) = \mathbf{r}(t)+ \sqrt{2D\Delta t} \xi \mathbf{t}_{0} + \sqrt{2D\Delta t} \chi \mathbf{t}_{1} 
\end{equation}
where $\mathbf{t}_{0}$, $\mathbf{t}_{1}$ furnish an orthonormal basis of the local tangent space and $\xi$ and $\chi$ refer to samples from a Gaussian density with vanishing mean and variance equal to unity. Due to the local nature of the tangent space, however, the updated position may violate the condition specified by Eq.~(\ref{implicit_surface}) and hence, it needs to be corrected according to
\cite{holyst1999diffusion}
\begin{equation}\label{surface_projection}
\mathbf{r}(t+\Delta t) = \mathbf{r}^{\ast}(t+\Delta t) - \frac{\psi(\mathbf{r}^{\ast}(t+\Delta t))\nabla\psi(\mathbf{r}^{\ast}(t+\Delta t))}{\vert\nabla \psi(\mathbf{r}^{\ast}(t+\Delta t))\vert^{2}}. 
\end{equation}
$\mathbf{r}(t+\Delta t)$ represents the particle's updated position on the surface after one completed simulation time step $\Delta t$.

Clearly, metaball models of surfaces (Eq.~(\ref{blobby_blend})) satisfy the condition given by Eq.~(\ref{implicit_surface}). In addition to the in Sec.~\ref{comp_model_geo} already mentioned major advantages they offer, their defining properties also make them ideally suited to efficiently construct the local differential geometry necessary to implement the described free diffusion algorithm. In particular, the all-important surface gradient and surface unit normal $\mathbf{n}(\mathbf{r})= \nabla\phi(\mathbf{r}) / \vert\nabla\phi(\mathbf{r}) \vert$ can be calculated swiftly even for more complex surfaces. Because such surfaces are created by superposition (Eq.~(\ref{blobby_blend})) of simple local potentials (Eq.~(\ref{potential})),  one can exploit the exact analytical form of the surface gradient. Thus, the need for slow numerical differentiation can be circumvented. In turn, the surface normal suffices to construct an orthonormal basis \cite{hughes1999building, frisvad2012building} and hence, the local tangent plane (Fig.~\ref{fig:local_frame}) that is required in Eq.~(\ref{free_membrane_propagation}).

 \subsection{Bimolecular reactions on single curved membranes}
\label{bimol_reactions_mem}
Similarly to the free diffusion case, by exploiting the surface's locally Euclidean geometry, the components of stochastic algorithms that describe bimolecular reactions on flat 2d membranes can be extended to metaball-based models of single curved membranes \cite{prustel2020stochastic}. 

The Green's function enhanced algorithms that implement the SCK model of bimolecular reactions, such as \cite{van2005simulating, oppelstrup2009first, johnson2014free, sokolowski2019egfrd, prustel2021space}, have in common that they divide a biochemical many-molecule system into a collection of isolated one- and two-particle problems. This division is justified when at most two molecules may encounter each other, at least on average, during the simulation time step. It is motivated by the availability of analytical expressions for the two-particle Green's functions. These expressions, in turn, are invoked to more efficiently propagate the encounter pairs that do not react during the simulation time step. Because the time evolution of isolated one-particle systems is governed by free diffusion, (Eq.~(\ref{free_membrane_propagation})), the remaining task at hand is to extend the pair-propagation mechanism, originally formulated for globally Euclidean spaces, to curved surfaces.
 
The two-body Green's function models an isolated pair of molecules diffusing on a Euclidean plane as two 2d disk-shaped particles $A$ and $B$ with diffusion constants $D_{A}$ and $D_{B}$ and encounter radii $a_{A}$ and $a_{B}$, respectively (Fig.~\ref{sub_fig:sck_two_step}) \cite{prustel2012exact}. The particles may form a bound state when their separation equals the encounter distance $a_{\text{enc}}= a_{A} + a_{B}$. In the bound state, the molecules may dissociate to form an unbound pair $A + B$. Such a two-body system is equivalent to two one-particle systems: One corresponds to a freely diffusing particle, while the other represents a particle diffusing with diffusion constant $D = D_{A} + D_{B}$ around a static, 'reactive' disk with radius $a_{\text{enc}}$. Importantly, the single particle's motion in the presence of a static disc is not described by Eq.~(\ref{free_membrane_propagation}).  

More formally, this can be seen as follows. The two-particle diffusion equation that governs the time evolution of the Green's function 
\begin{eqnarray}\label{2bodySmol}
&&\frac{\partial}{\partial t}p(\mathbf{r}_{A},  \mathbf{r}_{B}, t \vert \mathbf{r}_{A0},  \mathbf{r}_{B0}) = \nonumber\\
&&(D_{A}\nabla_{\mathbf{r}_{A}}^{2}+D_{B}\nabla_{\mathbf{r}_{B}}^{2}) p(\mathbf{r}_{A},  \mathbf{r}_{B}, t \vert \mathbf{r}_{A0},  \mathbf{r}_{B0}), \quad \vert\mathbf{r}_{B} - \mathbf{r}_{A}\vert \geq a_{\text{enc}},
\end{eqnarray}
can be rewritten \cite{sokolowski2019egfrd} by introducing center-of-diffusion coordinates
\begin{eqnarray}
\mathbf{R}&=&\frac{D_{B}}{D} \mathbf{r}_{A}+\frac{D_{A}}{D} \mathbf{r}_{B}\label{Rcoo}, \\
\mathbf{r}&=&\mathbf{r}_{B}-\mathbf{r}_{A}\label{relCoo}
\end{eqnarray}
as
\begin{equation}\label{separated-two-body-DE}
\frac{\partial}{\partial t} p(\mathbf{R}, \mathbf{r}, t \vert \mathbf{R}_{0},  \mathbf{r}_{0}) = 
(D_{R}\nabla_{\mathbf{R}}^{2} + D\nabla_{\mathbf{r}}^{2})p(\mathbf{R},  \mathbf{r}, t \vert \mathbf{R}_{0},  \mathbf{r}_{0}), \quad\vert\mathbf{r}\vert \geq a_{\text{enc}},
\end{equation}
where 
\begin{eqnarray}
D&=& D_{A}+D_{B}, \\
D_{R} &=& \frac{D_{A}D_{B}}{D}.
\end{eqnarray}
The ansatz $p(\mathbf{R},  \mathbf{r}, t \vert \mathbf{R}_{0},  \mathbf{r}_{0})= p(\mathbf{R}, t\vert \mathbf{R}_{0})p(\mathbf{r}, t\vert \mathbf{r}_{0})$ leads to \cite{sokolowski2019egfrd}
\begin{eqnarray}
\frac{\partial}{\partial t}p(\mathbf{R}, t\vert \mathbf{R}_{0}) &=& D_{\text{R}}\nabla^{2}_{\mathbf{R}} p(\mathbf{R}, t\vert \mathbf{R}_{0}),\label{center_eq}\\
\frac{\partial}{\partial t}p(\mathbf{r}, t\vert \mathbf{r}_{0}) &=& D \nabla^{2}_{\mathbf{r}} p(\mathbf{r}, t\vert \mathbf{r}_{0}), \quad\vert\mathbf{r}\vert \geq a_{\text{enc}}.\label{smol_bc}
\end{eqnarray}
Thus, as previously alluded to, one finds that the two-body diffusion equation (Eq.~(\ref{2bodySmol})) is equivalent to two independent stochastic processes: The center of diffusion $\mathbf{R}$ undergoes free diffusion in 2d, while the non-trivial part of the bimolecular reaction dynamics of an isolated pair is described by the Green's function of the interparticle vector $\mathbf{r}$ (Eq.~(\ref{smol_bc})). Because $p(\mathbf{r}, t\vert \mathbf{r}_{0})$ is only defined for $\vert\mathbf{r}\vert \geq a_{\text{enc}}$,
one has to impose a boundary condition at the encounter distance $\vert\mathbf{r}\vert = a_{\text{enc}}$ that specifies the type of bimolecular reaction.  

Reversible reactions can be introduced by the backreaction boundary condition \cite{agmon1984diffusion, kim1999exact, prustel2012exact, agmon1990theory}. Using the 2d polar coordinate representation of the position vector $\mathbf{r} = (r, \theta)$ and $2 \pi p(r, t\vert r_{0}) = \int^{2\pi}_{0} d \theta \, p(\mathbf{r}, t\vert \mathbf{r}_{0})$,
it reads as
\begin{equation}
\label{backreaction_bc}
2 \pi a_{\text{enc}}D\frac{\partial p(r,t\vert r_{0})}{\partial r}\bigg\vert_{r=a} = \kappa_{a}p(a,t\vert r_{0}) - \kappa_{d}[1-S(t\vert r_{0})],
\end{equation}
Here, $S(t\vert r_{0}) = 2 \pi \int^{\infty}_{a_{\text{enc}}}p(r, t\vert r_{0})dr$ is the survival, or separation probability of a pair of molecules that are initially at $t = 0$ separated by $r_{0}$ \cite{agmon1990theory}. 
The parameters $\kappa_{a}$ and $\kappa_{d}$ are referred to as intrinsic association and dissociation rate, respectively. They have to be distinguished from the macroscopic on- and off-reaction rates $k_{\text{on}}$, $k_{\text{off}}$ that appear in mass action rate equations \cite{shoup1982role, gosele1984reaction, rice1985diffusion, agmon1990theory, yogurtcu2015theory}.

In the limit $\kappa_{d}\rightarrow\infty$, the backreaction boundary condition reduces to the Collins-Kimball, or radiation boundary condition \cite{collins1949diffusion, agmon1990theory} that describes irreversible associations. In the limit $\kappa_{a},\rightarrow\infty$ it further reduces to the classical Smoluchowski, or totally absorbing, boundary condition \cite{smoluchowski1917mathematical}  
\begin{equation}\label{g_abs_bc}
p( r = a, t\vert r_{0}) = 0,
\end{equation}
that describes a pair instantaneously reacting upon encounter. 

Now, the analytical form of the Green's function that solves the boundary-value problem (Eq.~(\ref{smol_bc}) and either Eq.~(\ref{backreaction_bc}) or one of its limits) is used in a simulation to update the position vector $\mathbf{r}$ that must satisfy the constraint $\vert\mathbf{r}\vert \geq a_{\text{enc}}$. Because the expression of the Green's function is given in terms of polar coordinates $\mathbf{r}=(r, \theta)$, the position update involves two steps. First, one draws a new distance $r$. Second, one samples the angle $\theta$ between the position vectors $\mathbf{r}_{0}$, $\mathbf{r}$ before and after the update, respectively, given that the new distance is $r=\vert\mathbf{r}\vert$. In the case of a curved surface, this procedure cannot be performed with respect to the global standard 3d coordinate space furnished by the canonical basis vectors $\mathbf{e}_{1}=(1,0,0)$, $\mathbf{e}_{2}=(0,1,0)$, $\mathbf{e}_{3}=(0,0,1)$. Instead, as a prerequisite, $\mathbf{r}_{0}$ has to be transformed into the 2d coordinate space of the local tangent space. After sampling the new position $\mathbf{r}$ in the local tangent plane's coordinate space, $\mathbf{r}$ is transformed back into the standard 3d coordinate system. Finally, the updated $\mathbf{r}$ and $\mathbf{R}$ give rise to the new positions $\mathbf{r}_{A}$, $\mathbf{r}_{B}$ of the encounter pair. 

By contrast, while free diffusion updates require the local tangent plane as well, as is evident from (Eq.~(\ref{free_membrane_propagation})), there is no need to transform the involved position vectors into the coordinate system furnished by the basis vectors of the local tangent plane.
\subsection{Diffusion and bimolecular reactions at cell-cell contacts}
Turning to cellular signaling processes that involve sites of cell-cell contact, these algorithms have to be extended further to capture the novel features that arise from these special morphologies. For instance, the sheer presence of sites of cell-cell contact imposes constraints even on free membrane diffusion. If the ectodomain of a membrane complex is too long to physically fit in the intermembrane gap, its diffusive motion into the site of contact is obstructed. To capture this behavior in the simulation, one can assign a 'steric height' $h_{\text{steric}} > 0$ to membrane complexes. Then, the algorithm rejects those diffusion steps of membrane complexes with $h_{\text{steric}} > 0$ that would result in updated positions (Eq.~(\ref{free_membrane_propagation})) whose distances to the apposing membrane are incompatible with $h_{\text{steric}}$. Conversely, intercellular bound complexes spanning the cell-cell contact gap cannot assume new positions in conflict with another length, the 'encounter height' $h_{\text{enc}}$ that we will detail below. Accordingly, those propagation updates are rejected as well. In this fashion, free membrane diffusion alone can result in local depletion of some complexes and accumulation of others in the presence of contact sites, thereby reshaping the local spatial distribution of membrane proteins. Note that by employing the metaballs' properties (Eq.~(\ref{sign_check})) the checks whether to accept or reject a diffusion update are efficient and straightforward (Fig.~\ref{sub_fig:constrained_prop}).
\begin{figure}
\includegraphics[scale=0.4]{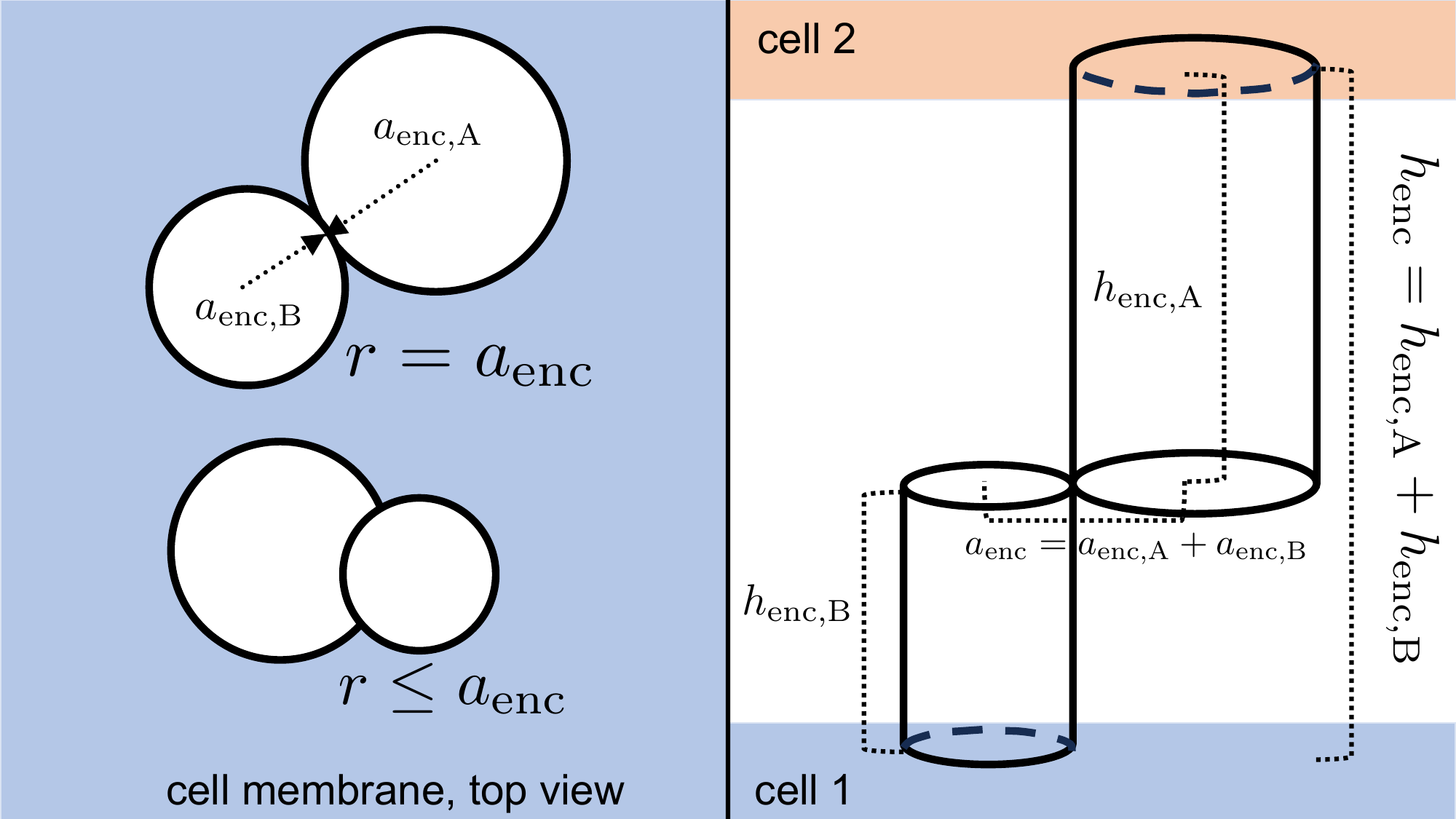}
\caption{
Encounter radius and height.
\newline
\textbf{Left:}
In the presence of a single membrane, the Smoluchowski picture models a pair of molecular complexes $A, B$ as 2d discs characterized by their encounter radius $a_{\text{enc,A}}, a_{\text{enc,B}}$, respectively. Bimolecular reactions $A+B\rightarrow C$ are incorporated via boundary conditions imposed at the encounter radius $r = a_{\text{enc}} = a_{\text{enc,A}} + a_{\text{enc,B}}$ (Eq.~(\ref{backreaction_bc})). In a simulation, the condition $r \leq a_{\text{enc}}$ identifies encounter pairs.  
\textbf{Right:}
By contrast, a cell-cell contact creates the possibility of intercellular encounter pairs. Then, 2d discs cease to model such pairs of complexes adequately, which instead may be more accurately depicted by 3d cylinders. In addition to an encounter radius $a_{\text{enc}}$, cylinders are characterized by an encounter height $h_{\text{enc}}$. Encounter pairs are now identified by two conditions, the usual one involving encounter radii and a novel one that requires the compatibility of intermembrane gap distance and the combined encounter height $h_{\text{enc}}= h_{\text{enc,A}}+h_{\text{enc,B}}$ of the pair.     
}\label{fig:discs_to_cylinders}
\end{figure}
\begin{figure}
\includegraphics[scale=0.4]{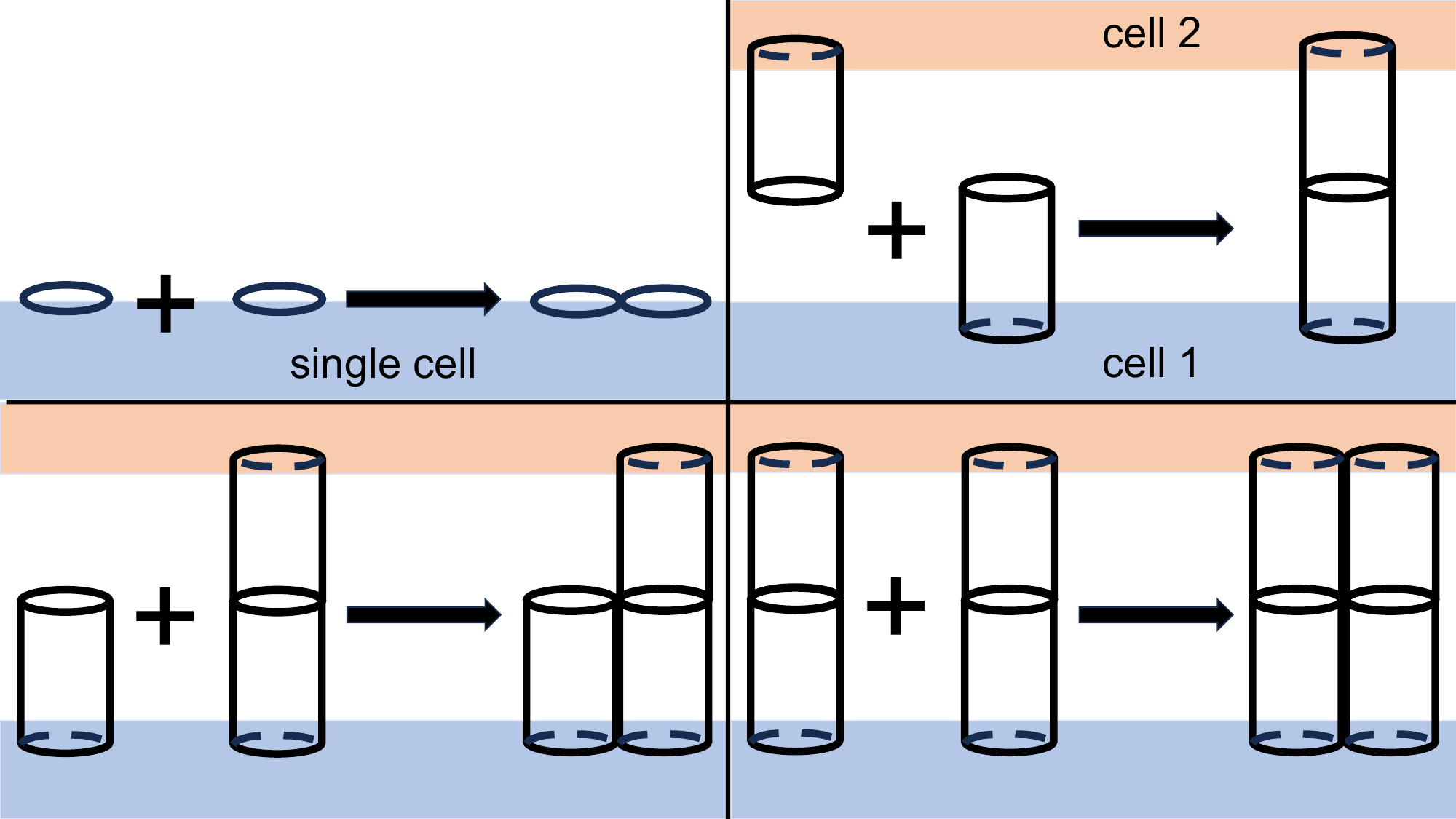}
\caption{
Single-membrane and intermembrane bimolecular reactions.
\newline 
\textbf{Top left:}
On a single membrane, particles that can undergo a bimolecular reaction may be depicted as 2d discs. \textbf{Top right to bottom right:} In the presence of a cell-cell contact, novel classes of bimolecular reaction possibilities emerge. Already in a concentration-based deterministic approach, the transition from a non-spatial to spatial description results in a substantially increased number of interaction possibilities \cite{angermann2012computational}. In a particle-based approach, the knowledge of the precise location of membrane complexes makes it possible to naturally enforce constraints that are based on spatial proximity, such as compatibility of intermembrane gap distance and encounter height. Particles are then more accurately modeled as 3d cylinders rather than as 2d discs.   
}\label{fig:intercellular_possibilities}
\end{figure}

We have seen in Sec.~\ref{bimol_reactions_mem} that the SCK model accounts for bimolecular associations by depicting membrane complexes as 2d discs characterized by their diffusion coefficients and encounter radii and that this picture is still  appropriate when the associations take place place on a curved membrane rather than on a flat 2d plane. However, cell-cell contact sites create the possibility of new types of bimolecular membrane associations (Fig.~\ref{fig:intercellular_possibilities}) that have no analogue in single membrane systems. In particular, when it comes to intercellular associations $A+B \rightarrow C$ (Fig.~\ref{fig:intercellular_possibilities}, top right) between particles that reside on one cell membrane and particles located on an apposing membrane, the depiction of membrane complexes as 2d discs becomes woefully inadequate. Instead, membrane complexes that can form intercellular bonds are more accurately described as 3d cylinders with (encounter) radii $a_{\text{enc}}$ and encounter heights $h_{\text{enc}}$ (Fig.~\ref{fig:discs_to_cylinders}). Now, there are then two encounter conditions to check in a simulation context: Besides the usual overlap criteria given by $r \leq a_{\text{enc}}$, the encounter height $h_{\text{enc}}=h_{\text{enc, A}}+h_{\text{enc, B}}$ has to be larger than or or equal to the gap distance. Note that here the encounter radius overlap criteria involves the 2d local tangent plane projection of $r=\vert\mathbf{r}_{B}-\mathbf{r}_{A}\vert$ and not its full 3d counterpart. Importantly, both conditions need to be simultaneously fulfilled to conclude that, indeed, the two membrane complexes encountered each other. The $h_{\text{enc}}$ related condition can only be checked by direct overlap; there is no analytical expression of an Green's function derived encounter probability available that could complement the direct overlap check. In particular, in this context, one cannot replace the Greens' function of a point-like particle diffusing around a 2d disc by the one that analogously describes the diffusion of a such a particle around a 3d cylinder. The reason is that the stochastic motion along the height of the cylinder is not due to diffusion in 3d space, but it depends on both the 2d diffusion of the complexes on their respective membranes and the curvature of their local neighborhood membrane portions.  

Finally, we point out that the sampling procedure of the interparticle vector described in Sec.~\ref{bimol_reactions_mem} becomes more involved for intermembrane associations. Now, one has to project the position of one complex of the encounter pair into the local tangent space of its partner complex on the apposing membrane, execute the propagation of the interparticle and center-of-mass vector with respect to the tangent plane's coordinate system as discussed before and finally project the appropriate one of the two updated positions back on the other membrane.
\section{Applications}
\label{applications}
\subsection{Numerical construction of the Green's function of a reversibly reacting isolated pair on a sphere}
\label{numeric_GF_app}
First, we assessed the algorithm's accuracy for diffusion-influenced bimolecular chemical reactions on the level of individual particles that diffuse on the same single membrane.  
To this end we simulated the irreversible as well as reversible reaction of an isolated molecule pair diffusing on the surface of a sphere. Thus, we could numerically construct the corresponding probability density function, that is, the Green's function $p(\theta, t\vert \theta_{0})$ (Eq.~(\ref{p_abs_sphere})), where $\theta$ refers to the polar angle of standard spherical coordinates, and compare it with the known exact analytical form of the Green's function \cite{grebenkov2019reversible}.
\begin{figure}
\includegraphics[scale=0.43]{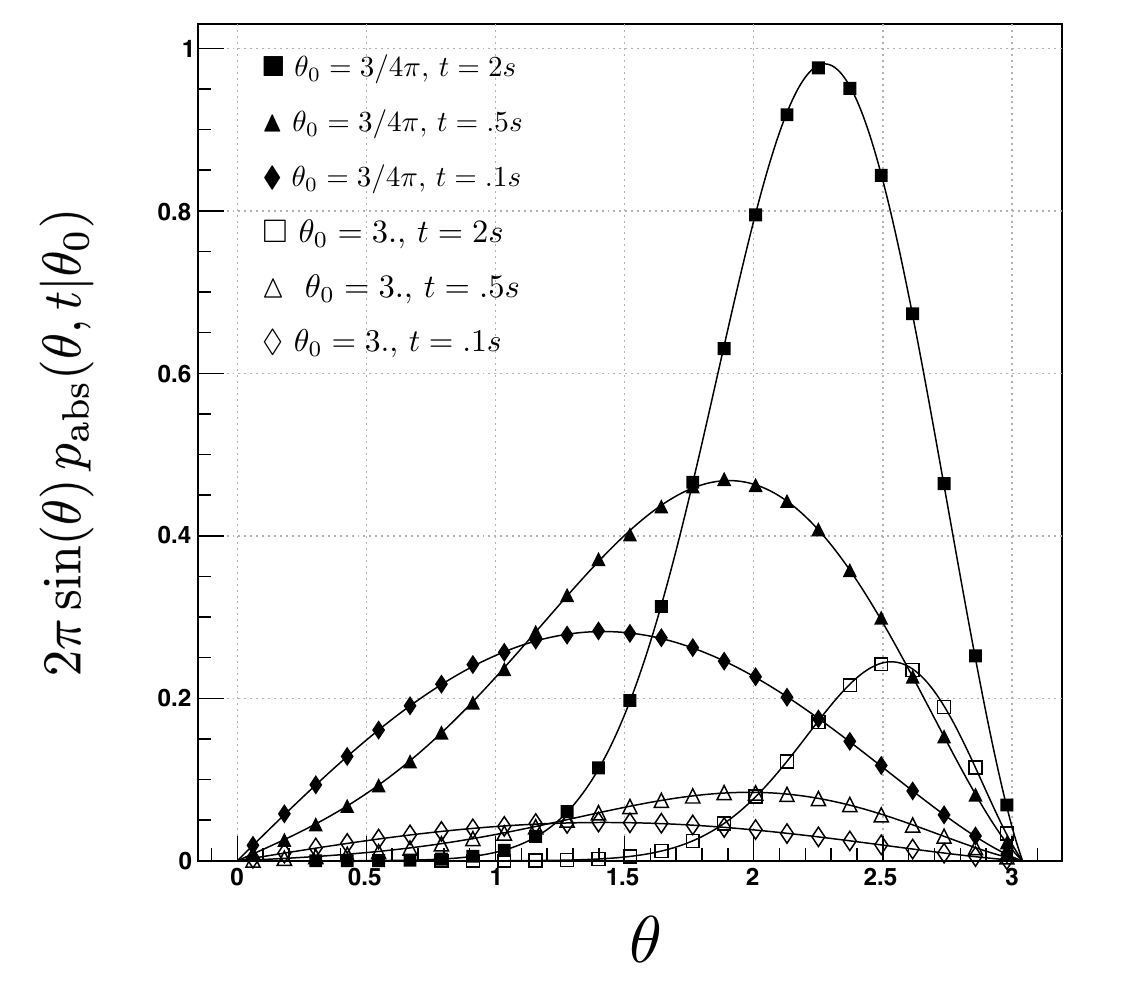}
\caption{
Numerical construction of the GF $p_{\text{abs}}(\theta, t\vert \theta_{0})$ describing the diffusion-controlled irreversible binding of an isolated molecule pair on the surface of a sphere.
The figure shows the $\theta$ dependence of $p_{\text{abs}}(\theta, t\vert \theta_{0})$ at different times and for two different values of the initial position $\theta_{0}$, as indicated. 
The solid lines correspond to the exact analytical expressions (Eq.~(\ref{p_abs_sphere})).
The various markers correspond to the normalized histograms of the molecule's simulated final position at time $t$.
}\label{fig:p_sphere_abs}
\end{figure}

The simulation setup involved an isolated pair of molecules $A$ and $B$. Molecule $A$ was held fixed at the sphere's south pole $\theta = \pi$, while molecule $B$ was placed at an initial position $\theta_{0}$. During a simulation run, molecule $B$ underwent a diffusive motion and potentially associated with molecule $A$ that was either modeled as a perfect sink, partially-reactive sink or reversibly reacting target. In the case of a reversible reaction, a bound $B$ molecule could dissociate and subsequently bind again possibly many times. We also performed simulations for different times $t_{\text{sim}}$ to check the time dependence of the probability density function.
After each run, $B$'s final position was recorded, unless it was bound to $A$, and a new simulation was started.
The resulting histogram was normalized to take into account the number of bound states at $t_{\text{sim}}$. 
The radius of the sphere, the encounter radius, the diffusion constant and simulation time step were chosen as $R=1\mu\text{m}$, $a_{\text{enc}}=0.1\mu\text{m}$, $D=1\mu\text{m}^{2}\text{s}^{-1}$ and $\Delta t = 0.001\text{s}$, respectively. The other parameters,
the intrinsic association and dissociation rates $\kappa_{a}$ and $\kappa_{d}$ (Eq.~(\ref{backreaction_bc_sphere})), respectively, were varied.

\begin{figure}
\includegraphics[scale=0.325]{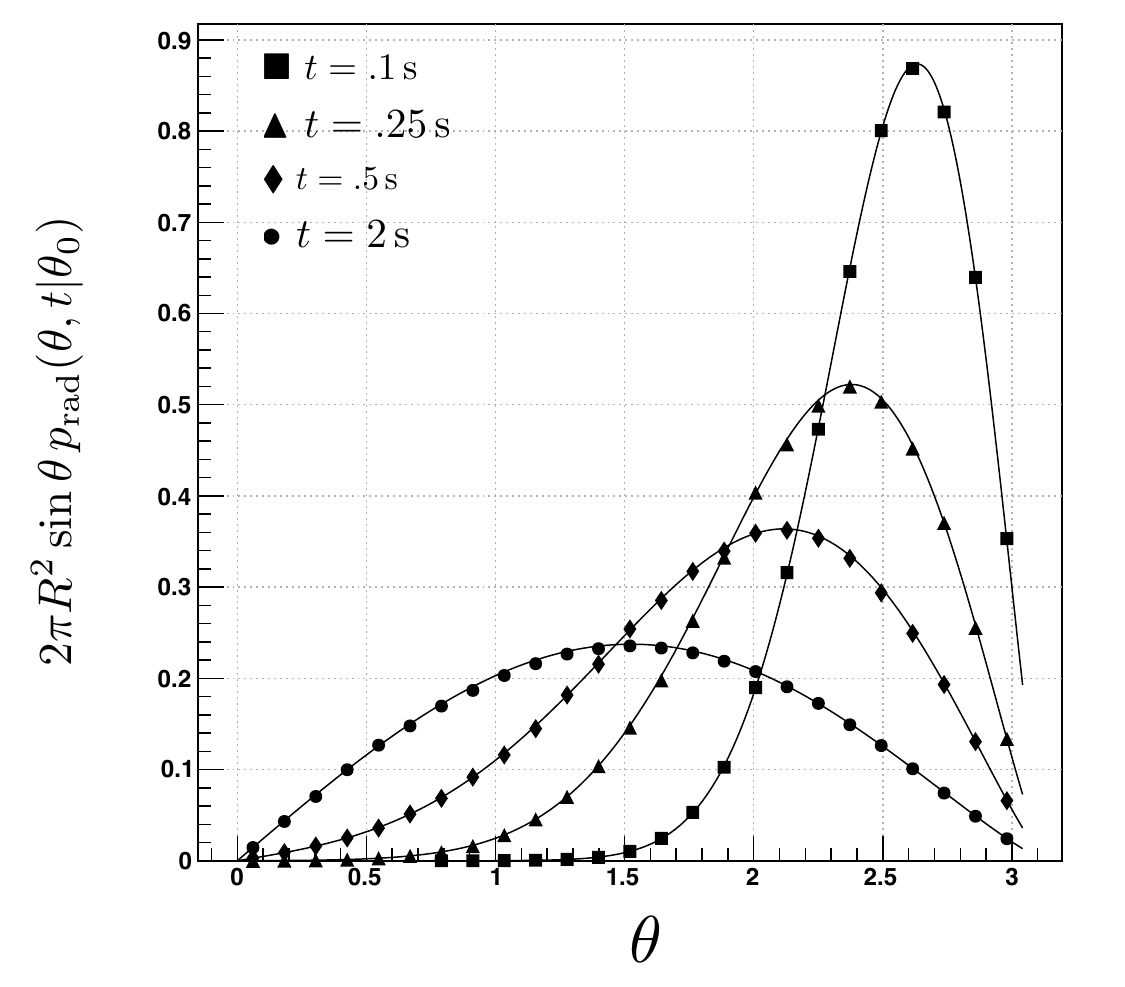}
\includegraphics[scale=0.325]{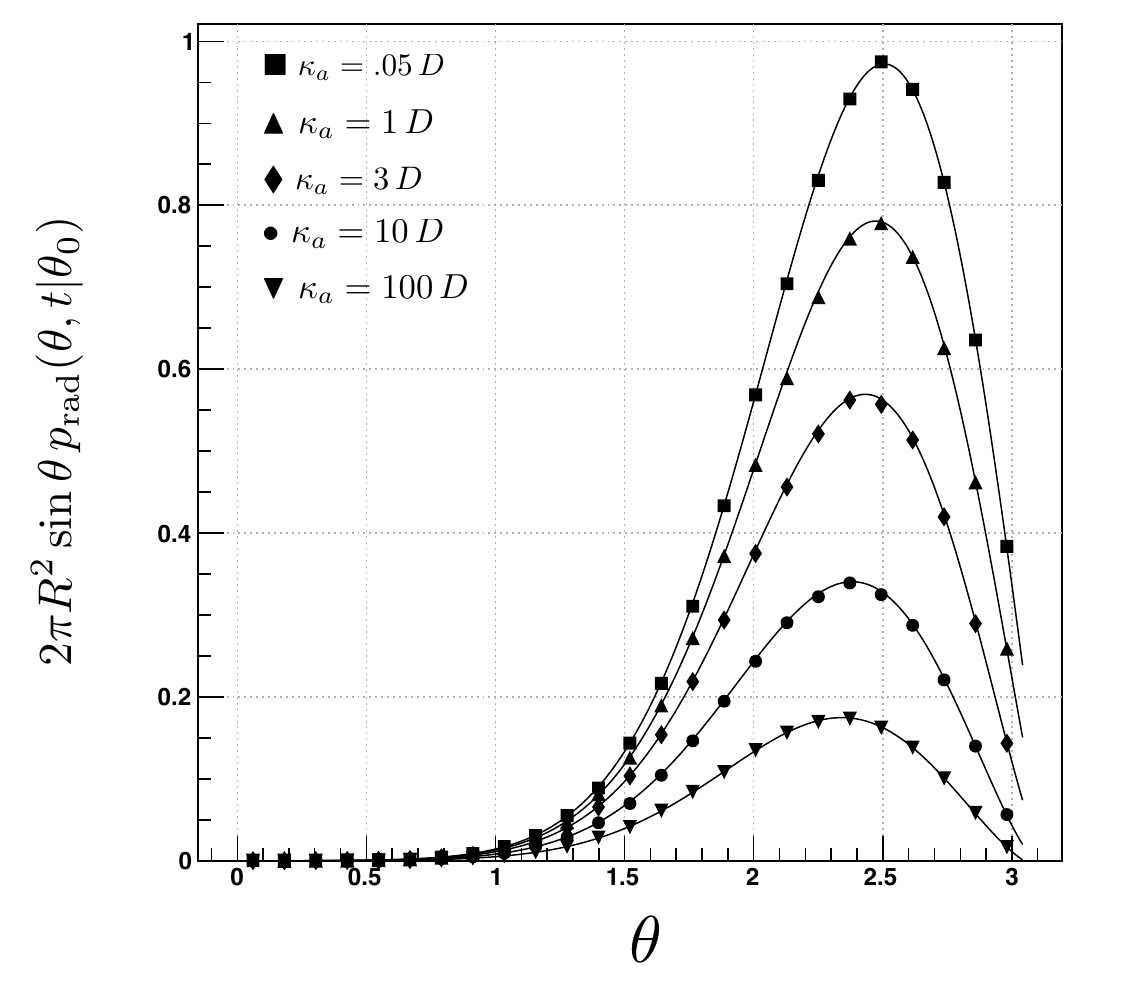}
\caption{
Numerical construction of the GF $p_{\text{rad}}(\theta, t\vert \theta_{0})$ describing the diffusion-influenced irreversible binding of an isolated molecule pair on the surface of a sphere.
\newline
\textbf{Left:} The figure shows the $\theta$ dependence of the GF satisfying a partially absorbing boundary condition ($\kappa_{a}=2\mu\text{m}^{2}\text{s}^{-1}$) at different times $t$, as indicated. Initially, the two molecules are at contact $\theta_{0}=\theta_{a}$, see Eq.~(\ref{theta_a_def}).
\textbf{Right:} The figure shows the $\theta$ dependence of $p_{\text{rad}}(\theta, t\vert \theta_{0})$ for different values of the intrinsic association constant $\kappa_{a}$, as indicated, at time $t=0.2\,\text{s}$. The initial position is $\theta_{0}=3$. 
The other parameters $D, R, a_{\text{enc}}$ are $D=1\mu\text{m}^{2}\text{s}^{-1}$, $R=1\mu\text{m}, a_{\text{enc}}=0.1\mu\text{m}$.
The solid lines are obtained from the exact analytical expressions given in \cite{grebenkov2019reversible}.
The various markers correspond to the normalized histograms of the molecule's simulated final position at time $t$.
}\label{fig:p_sphere_rad}
\end{figure}

\begin{figure}
\includegraphics[scale=0.325]{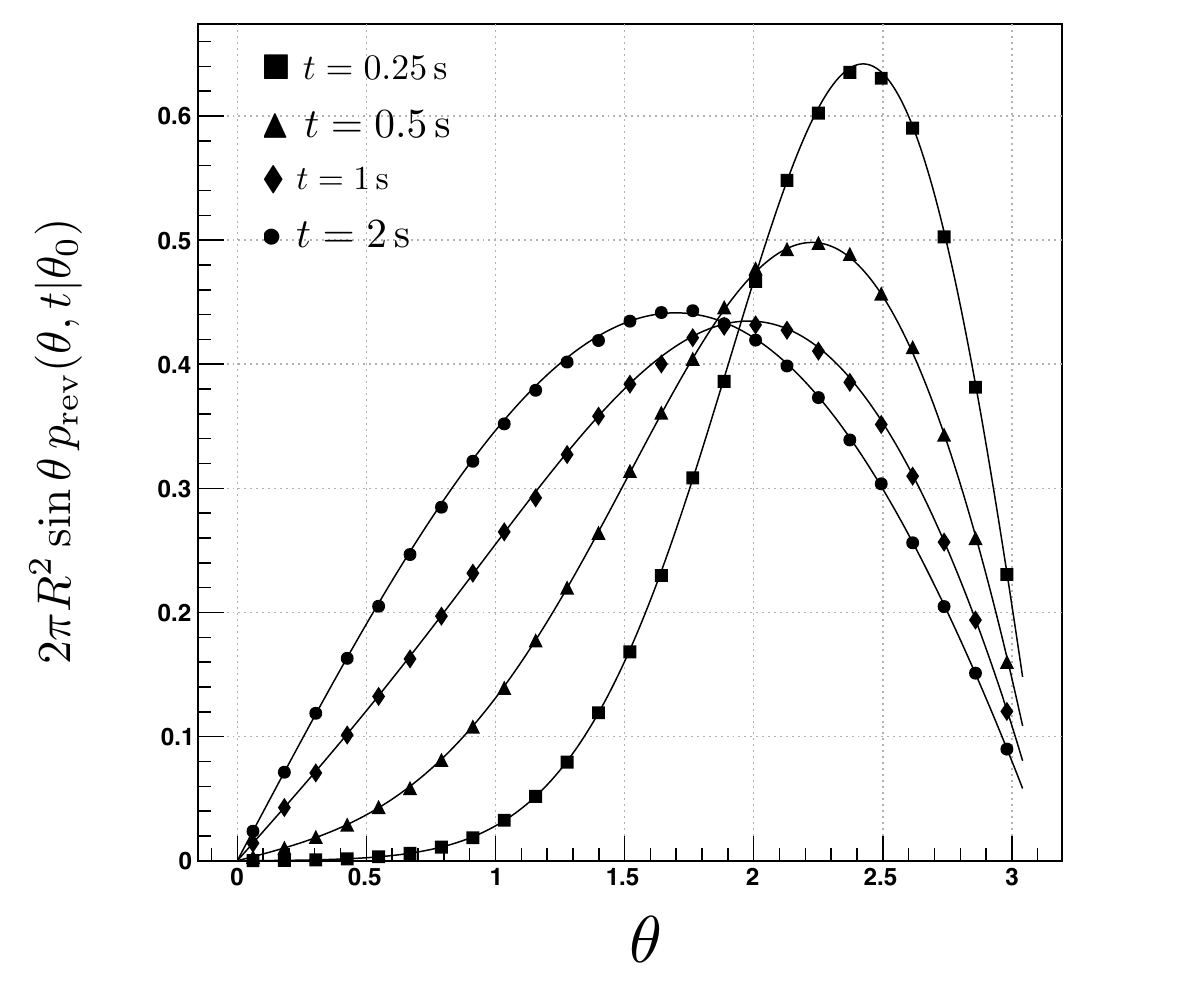}
\includegraphics[scale=0.325]{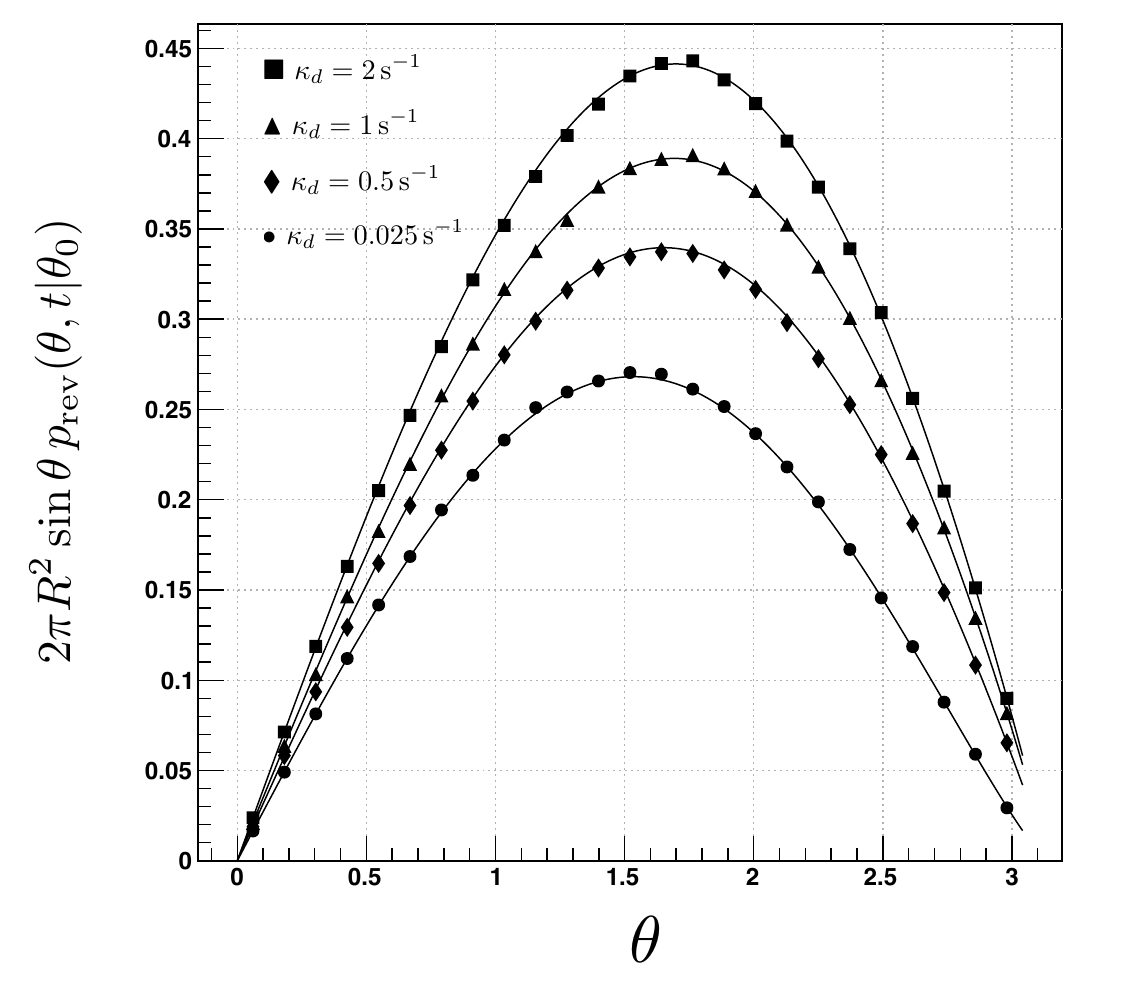}
\caption{
Numerical construction of the GF $p_{\text{rev}}(\theta, t\vert \theta_{0})$ describing the diffusion-influenced reversible binding of an isolated molecule pair on the surface of a sphere.
\newline
\textbf{Left:} The figure shows the $\theta$ dependence of the GF satisfying the reversible backreaction boundary condition \cite{grebenkov2019reversible} at different times $t$, as indicated. Additional parameters are $\theta_{0}=3$, $\kappa_{a}=2\mu\text{m}^{2}\text{s}^{-1}$ and $\kappa_{d}=2\,\text{s}^{-1}$.
\textbf{Right:} The figure shows the $\theta$ dependence of $p_{\text{rev}}(\theta, t\vert \theta_{0})$ for various values of the intrinsic dissociation constant $\kappa_{d}$, as indicated, at time $t=2\,\text{s}$. Further parameters are $\theta_{0}=3$ and $\kappa_{a}=2\mu\text{m}^{2}\text{s}^{-1}$. 
The remaining parameters $D, R, a_{\text{enc}}$ are the same as in Fig.~\ref{fig:p_sphere_rad}. The solid lines describe
the exact analytical expressions derived in \cite{grebenkov2019reversible}.
The various markers correspond to the normalized histograms of the molecule's simulated final position at time $t$.
}\label{fig:p_sphere_rev}
\end{figure}

The simulation results displayed in Figs.~\ref{fig:p_sphere_abs}, \ref{fig:p_sphere_rad} and \ref{fig:p_sphere_rev} show for all simulated cases excellent agreement with the theoretical prediction \cite{grebenkov2019reversible} (see also Sec.~\ref{appendix}). 
\begin{figure}
\includegraphics[scale=0.7]{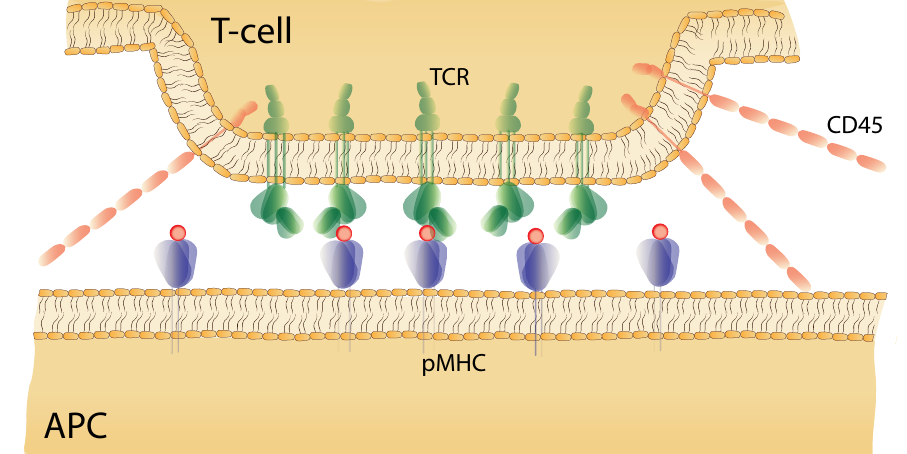}
\includegraphics[scale=0.285]{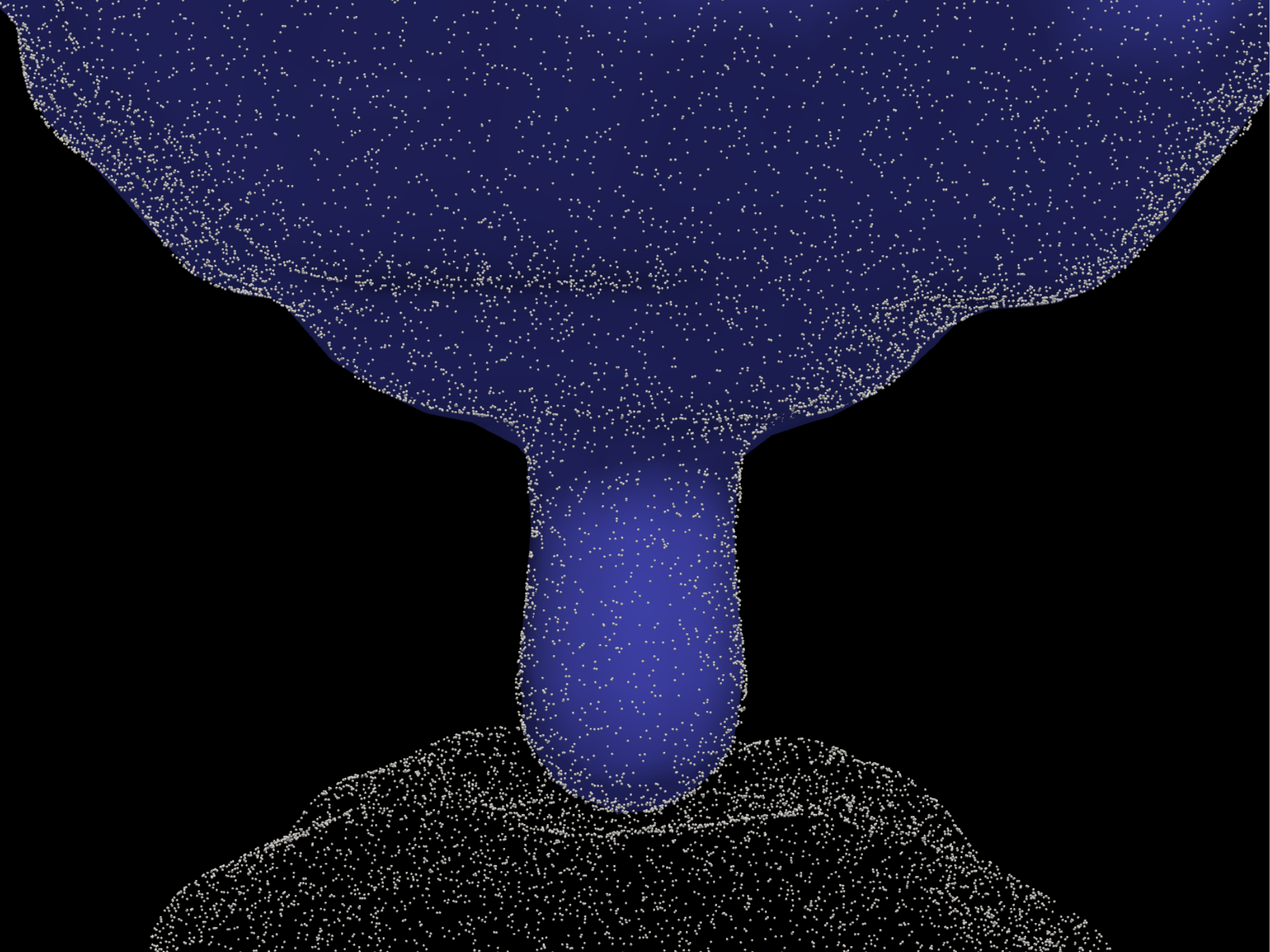}
\caption{
Simple model of T-cell receptor activation.
\newline
\textbf{Top:}
Schematic depiction of a previously considered model \cite{fernandes2019cell} asserting that TCR activation rests on two biophysical quantities. One is the TCRs residence time in cell-cell contact sites that are depleted of large phosphatases, such as CD45. The other is the size of the contacts that in turn is determined by the contact geometry and the height of the depleted phosphatases. Then, any TCR residing without interruption within the contact site for longer than a critical time is considered to be activated. 
\textbf{Bottom:} Snapshot of a corresponding simulation. The blue-colored model represents a T-cell, the positions of the TCR and pMHC complexes are rendered as white dots. The APC itself is not shown. The CD45 phosphatase is not directly modeled; its effect is accounted for by the size of the contact site that gives rise to the TCRs residence time.   
}\label{fig:TCR_APC_interface}
\end{figure}
\subsection{Simple model of early phase of T-cell receptor activation}
\label{TCR_app}
We now demonstrate the utility of the new simulation approach by exploring the effects of contact geometry and local biochemistry for the early phase of T-cell receptor activation. To this end, we consider a previously suggested model \cite{fernandes2019cell} that asserts that TCR activation is essentially regulated by the TCRs dwell-time in cell-cell contact sites that are depleted of large phosphatases, such as CD45. The dwell time, in turn, depends on interactions between the TCRs and peptide-MHC (pMHC) complexes and the size of the contact region. Any TCR that resides without interruption within the contact region for longer than a critical time is considered activated. This time was assumed to be equal to $2s$ by the authors of the original study.

The crucial quantity that can be derived from this model is the activation probability, which is the likelihood that at least one TCR has been activated during the duration of the cell-cell contact.
In Ref.~\cite{fernandes2019cell}, this quantity is derived by invoking two other quantities: First, the authors consider the Green's function that describes free and bound TCRs. They compute this quantity numerically on a 2d flat circular-shaped domain that can grow over time. Furthermore, they impose a totally absorbing and a reflective, no flux boundary condition for free and bound TCR, respectively. We note that due to the completely absorbing boundary condition, it is not possible to impose an initial condition where the TCR starts at the boundary directly, as it should. Therefore, the TCR must start already inside the domain and the numerical solution does depend on this somewhat arbitrary choice of the initial location. It is not immediately clear, to what extent alternative initial TCR positions would influence the subsequent results. 
Finally, to obtain the activation probability, another quantity, the time-dependent rate of TCR entry into the domain, is calculated. To this end, an approximate expression for the mean trapping time in the presence of a perfect trap on a spherical surface is used \cite{weaver1983diffusion}. While the assumptions that underlie this expression make sense in the construction of the activation probability given in Ref.~\cite{fernandes2019cell}, they are not compatible with the molecular events that actually take place at the contact site. In biological reality, these sites are not 'perfect sinks' for the TCRs, instead the TCRs may escape the contact region and subsequently re-enter it many times. 

These considerations prompted us to study the model (Fig.~\ref{fig:TCR_APC_interface}, top) in an alternative fashion, using our simulation approach capable of tracking the diffusive motion of individual TCRs on realistic membrane morphologies over time. The residence time frequency distribution of the TCRs is obtained as a direct readout of such a simulation. With the simulations, we intend to address two basic questions. First, in the absence of ligands, how does the TCR residence time distribution depend on the size of the contact site? Second, how does the presence of pMHCs affect the residence time distribution? How does it depend on the TCR-pMHC reactions rates and the pMHC density?

Fig.~\ref{fig:TCR_APC_interface} (bottom) shows the geometric model of the T-cell and APC being at contact via a T-cell protrusion. The total area of the T-cell and APC is $\sim 487\mu m^{2}$ and $\sim 360\mu m^{2}$, respectively. The intercellular gap distance is
$15 nm$. The contact site is defined as the portion of the protrusion's tip that is inaccessible for membrane particles with a certain steric height $h_{\text{steric}}$. Thus, varying $h_{\text{steric}}$ allows us to study different contact site areas. Note that the contact sites we consider are static, in contrast to \cite{fernandes2019cell}. The relevant observable is the residence times of individual TCRs within the contact area. Any time an individual TCR enters the contact site, the entry time is recorded, and any time a TCR leaves this area, its residence time within the site is recorded. 

\begin{figure*}[t !]
\centering
  
    \begin{subfigure}[t]{0.325\textwidth}
        \centering
        \includegraphics[scale=0.16]{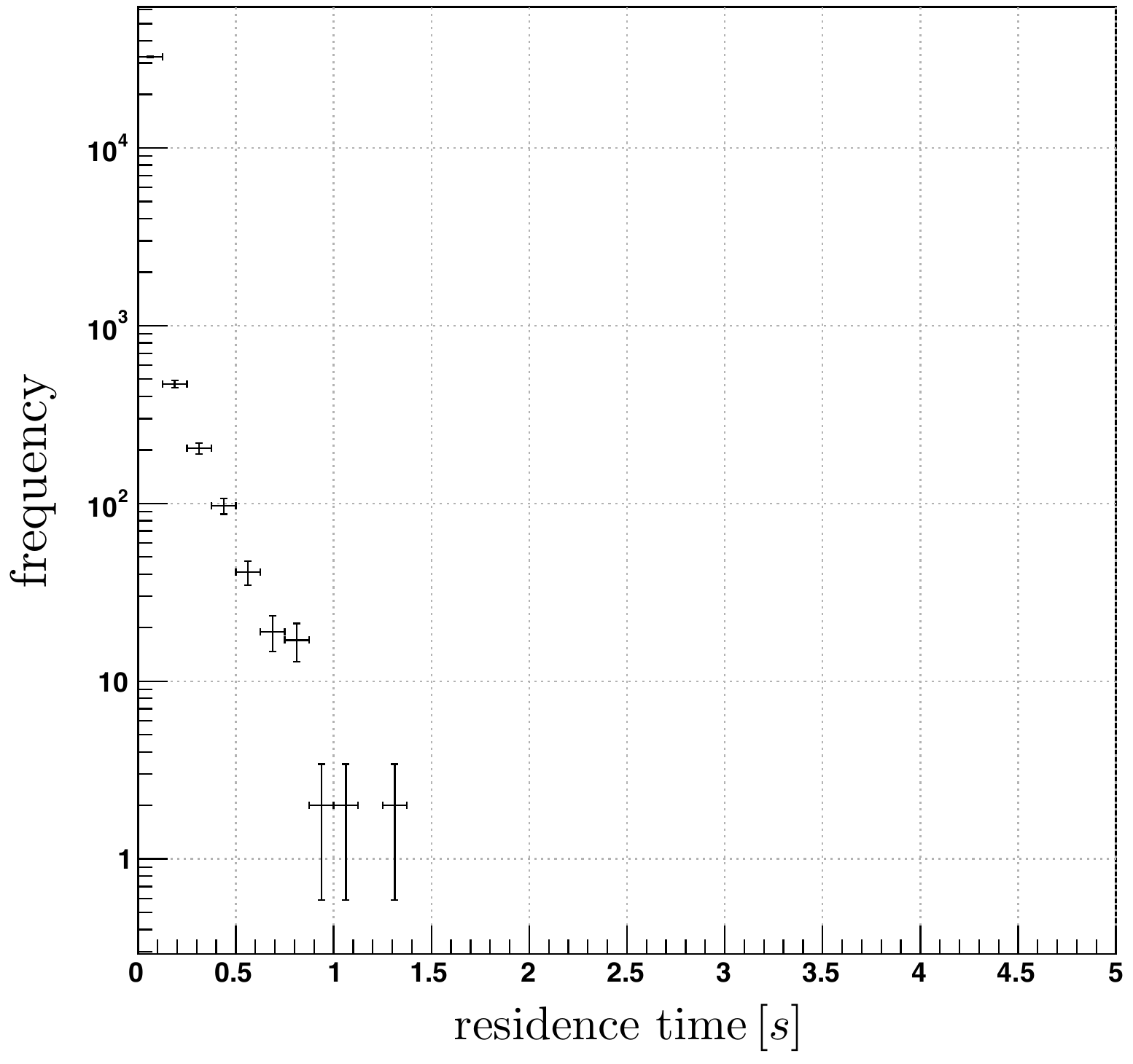}
        \caption{$r_{c} = 214nm$}
        \label{sub_fig:small_c}
    \end{subfigure}%
  ~
    \begin{subfigure}[t]{0.325\textwidth}
        \centering
        \includegraphics[scale=0.16]{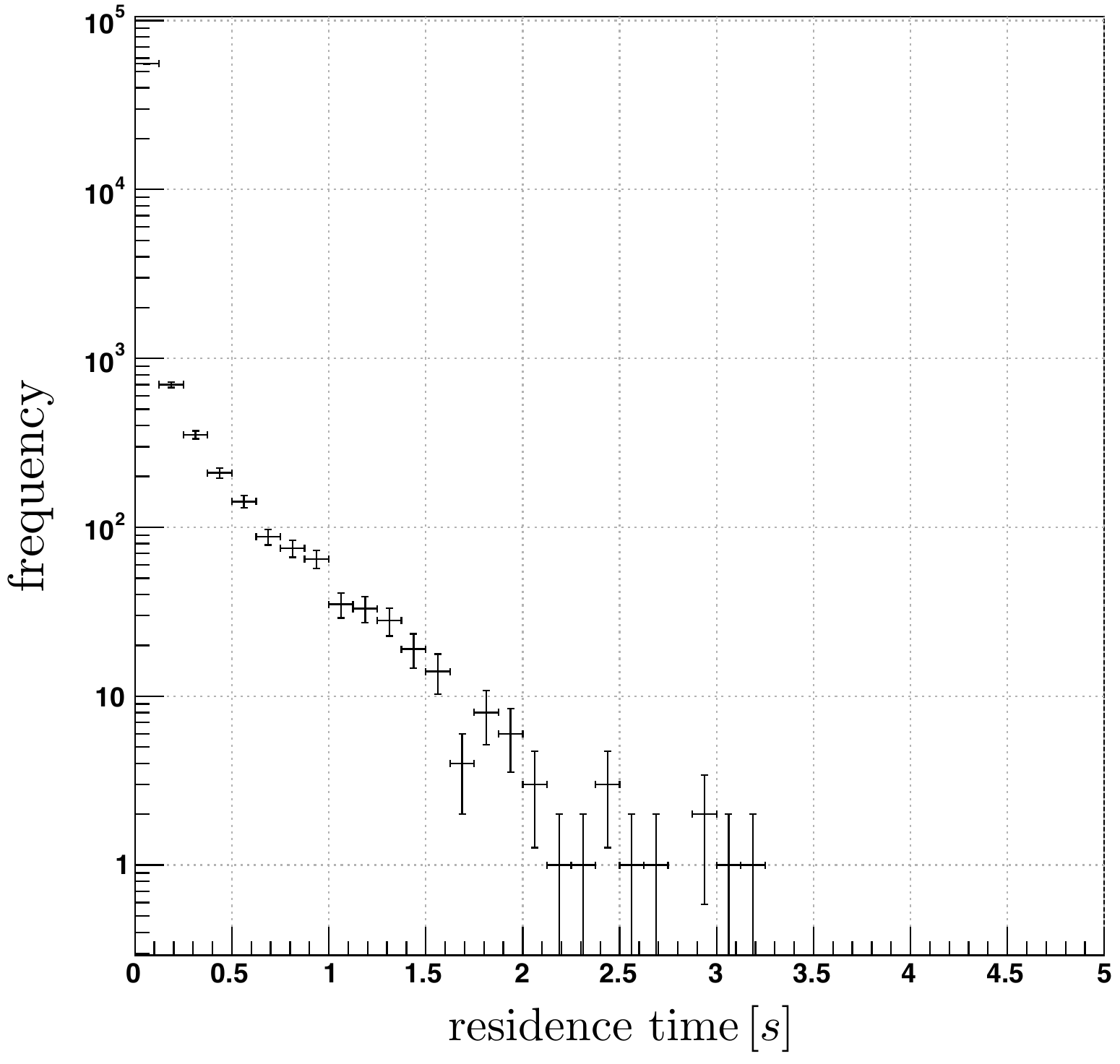}
        \caption{$r_{c}=346 nm$ }
        \label{sub_fig:medium_c}
    \end{subfigure}%
  ~
    \begin{subfigure}[t]{0.325\textwidth}
        \centering
        \includegraphics[scale=0.16]{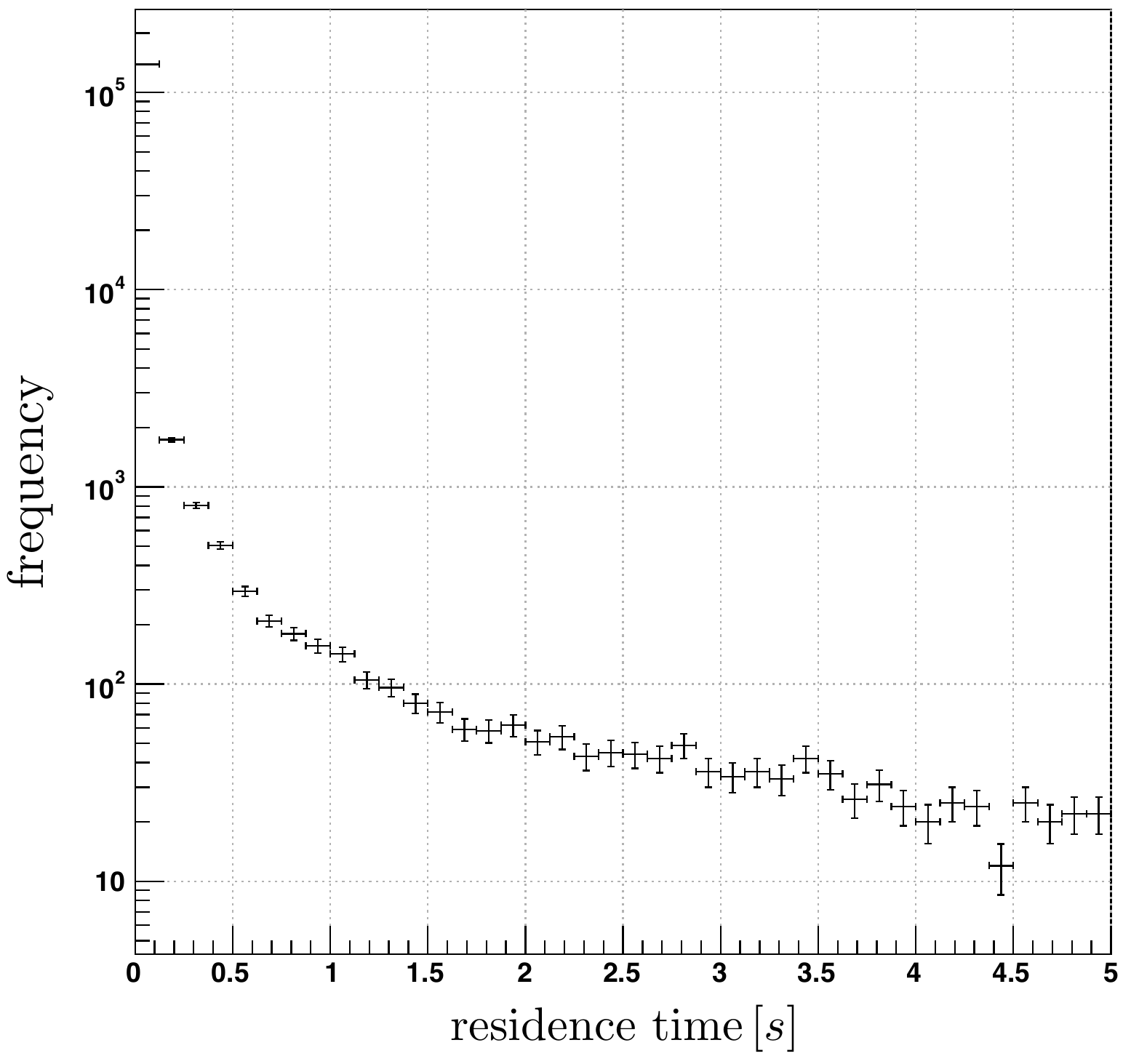}
        \caption{$r_{c}=1.05 \mu m$ }
        \label{sub_fig:large_c}
    \end{subfigure}%
    
\caption{
TCR residence time frequency distributions in the absence of pMHCs depending on different sizes of the contact site.
The parameters are $D_{\text{TCR}} = 0.05\mu m^{2}s^{-1}$, $[\text{TCR}]= 100\mu m^{-2}$.
}\label{fig:TCR_only}
\end{figure*}

We consider three different steric heights $h_{\text{steric}} = 25, 58, 750 nm$. The resulting contact site sizes are characterized by the radius $r_{c}=\sqrt{A_{c}/\pi}$, where $A_{c}$ denotes the contact site's area. Note that despite assigning a radius in this way, the contact site is not modeled by a flat circle.
The height $h_{\text{steric}} = 25 nm$ corresponds to the the lower, while $h_{\text{steric}} = 58 nm$ is somewhat larger than the upper limit of the range of ectodomain lengths that the differentially expressed CD45 isoforms display \cite{cordoba2013large, chen2021trapping}. Obviously, the height of $750 nm$ does not correspond to any membrane complex, but is chosen to create a large contact site area. Additional simulation parameters are the TCR's diffusion coefficient $D_{\text{TCR}}= 0.05 \mu m^{2} s^{-1}$, TCR membrane density $[\text{TCR}]=100\mu m^{-2}$ and contact duration $t_{sim} = 30s$. Initially, all TCRs are homogeneously distributed over the whole cells' area, including the contact region. 
The resulting TCR residence time frequency distributions are shown in Fig.~\ref{fig:TCR_only}.

Overall, the results of our detailed single-particle based stochastic simulations are compatible with the predictions made in Ref.~\cite{fernandes2019cell} using a simplified approach. Similar to those prior observations, we see a strong dependence of the residence time distribution on varying contact sizes, indicating a high sensitivity of the ligand-independent TCR activation probability towards changes in the contact radius $r_{c}$. In particular, the contact size that is experimentally observed during the T-cell scanning of APCs results in a vanishing probability that a TCR's residence time becomes larger than $2s$ (Fig.~\ref{sub_fig:small_c}). A modest increase of the radius to $r_{c}=346 nm$ is sufficient to considerably change the residence time distribution and to bring the system to a 'transition point', where at least a few copies of activated TCRs can arise (Fig.~\ref{sub_fig:medium_c}). Finally, increasing the size further, one finds a substantial amount of ligand-independent triggered TCRs (Fig.~\ref{sub_fig:large_c}).
\begin{figure*}[t !]
\centering  
    \begin{subfigure}[t]{0.325\textwidth}
        \centering
        \includegraphics[scale=0.16]{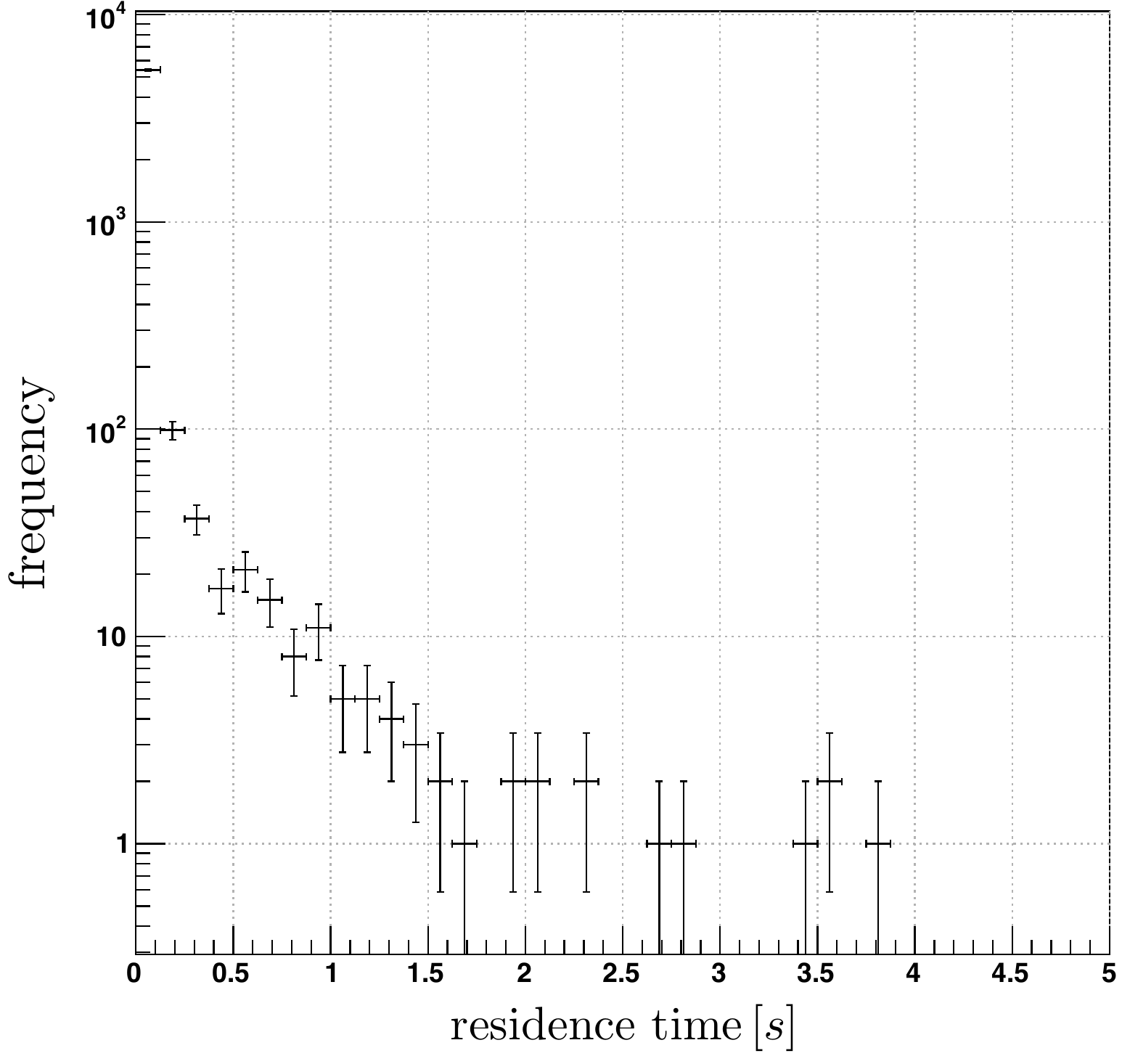}
        \caption{
        $[\text{pMHC}] = 30\mu m^{-2}$,\\
         $\kappa_{\text{off}} = 5s^{-1}$}
        \label{sub_fig:pMHC_30}
    \end{subfigure}%
  ~
      \begin{subfigure}[t]{0.325\textwidth}
        \centering
        \includegraphics[scale=0.16]{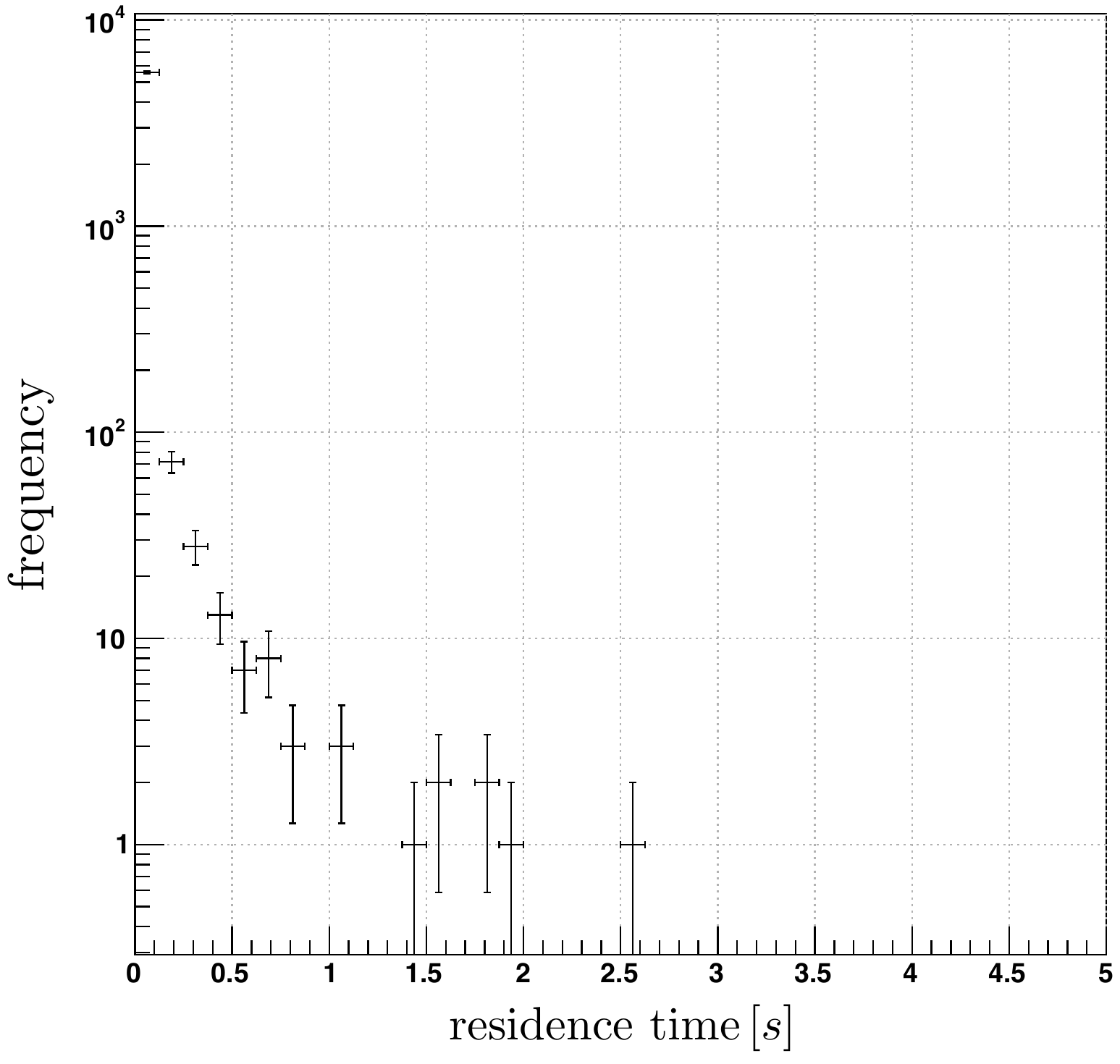}
        \caption{
        $[\text{pMHC}] = 3\mu m^{-2}$, \\
         $\kappa_{\text{off}} = 5s^{-1}$
        }
        \label{sub_fig:pMHC_3}
    \end{subfigure}%
  ~
    \begin{subfigure}[t]{0.325\textwidth}
        \centering
        \includegraphics[scale=0.16]{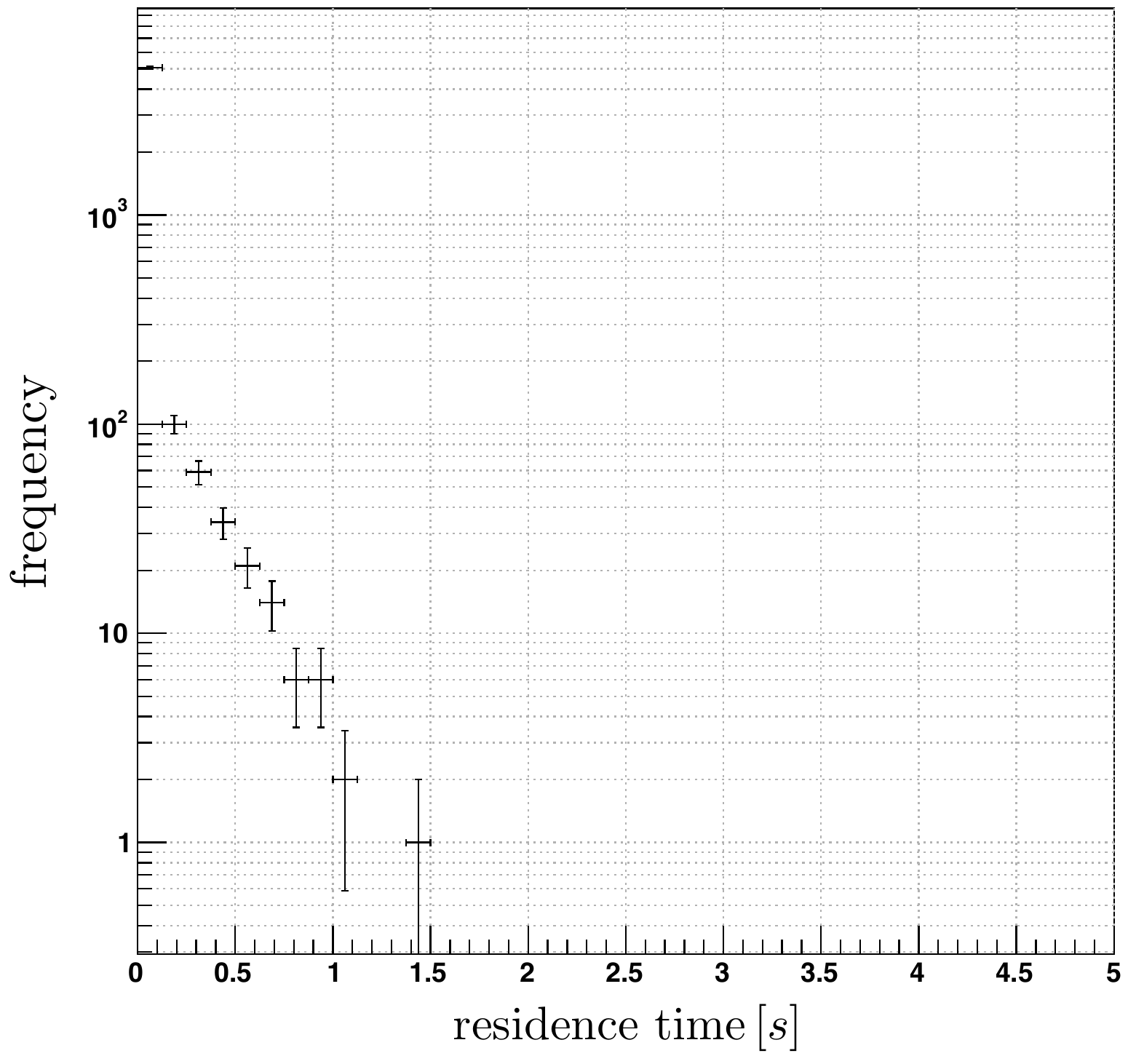}
        \caption{$[\text{pMHC}] = 60\mu m^{-2}$, \\
         $\kappa_{\text{off}} = 50s^{-1}$
         }
        \label{sub_fig:pMHC_60}
    \end{subfigure}%

\caption{
TCR residence time frequency distributions in the presence of pMHCs for different pMHC concentrations and $\kappa_{\text{off}}$ rates.
The other parameters are $D_{\text{TCR}} = D_{\text{pMHC}} = 0.05\mu m^{2} s^{-1}$, $[\text{TCR}]= 100\mu m^{-2}$, $r_{c} = 214nm$.
}\label{fig:TCR_pMHC}
\end{figure*}

Next, we explored the impact that the addition of pMHC ligands has on the TCR dwell-time distribution and hence, according to the model, on TCR activation probabilities. Here, we only consider a contact site radius of $r_{c} = 214nm$ that implies, as we have just seen, a virtually vanishing probability for ligand-dependent TCR activation. We further assume that $D_{\text{pMHC}} = D_{\text{TCR}} = 0.05\mu m^{2} s^{-1}$ and choose the on- and off-rates as $\kappa_{\text{on}} = 0.1 \mu m^{2} s^{-1}$ and $\kappa_{\text{off}} = 5 s^{-1}, 50 s^{-1}$, which correspond to a strong agonist and a weakly binding pMHC, respectively \cite{fernandes2019cell}.
Note that these reaction rates are the microscopic reaction rates of the SCK framework (Eq.~(\ref{backreaction_bc}). Their relation to the macroscopic rates that parametrize deterministic rate equations is discussed in Ref.~\cite{yogurtcu2015theory}. Importantly, $\kappa_{\text{off}}/\kappa_{\text{on}}$ gives the same $K_{d}$ value as the macroscopic rates. The simulated contact duration is now $t_{sim} = 5s$.
 
Figs.~\ref{sub_fig:pMHC_30} and \ref{sub_fig:pMHC_3} show the residence time distributions for two different densities of the strong agonist. Comparing Fig.~\ref{sub_fig:small_c} and \ref{sub_fig:pMHC_30} shows a substantial shift of the residence time distribution to larger dwell-times for low densities ($[\text{pMHC}] = 30\mu m^{-2}$) and still even for $10\times$ lower densities ($[\text{pMHC}] = 3\mu m^{-2}$, Fig.~\ref{sub_fig:pMHC_3}).  By contrast, the residence time distribution in the presence of weakly binding ligands is much less affected, although the density is now at least 2$\times$ higher ($ [\text{pMHC}] = 60\mu m^{-2}$, Fig.~\ref{sub_fig:pMHC_60}). 

In conclusion, several of the simulation results presented here are consistent with a sensitive and discriminatory TCR activation mechanism based on TCR residence times at cell-cell contacts, as claimed in Ref.~\cite{fernandes2019cell}. We also see that potential variations in the contact areas may negatively impact this system’s discriminatory power with regard to agonist and non-agonist peptide categories. A recent publication suggested a cell-biological mechanism that may, indeed, confine the contact area during T cell - APC interactions, thereby potentially protecting the discriminatory power of the local CD45 exclusion model \cite{jenkins2023antigen}. However, many questions remain unanswered at this point. For instance, it is unclear whether all 2d off-rates of TCR-peptide:MHC complexes that have been found to be non-stimulatory do, in fact, fall into the region that does not capture TCR longer than 2 seconds in the phosphatase depleted contact zone. Furthermore, it is still rather unclear which molecular mechanisms could provide a sharp cutoff of 2 seconds for successful activation of the TCR. Elucidating such questions will require detailed stochastic simulations, for which we are here providing an approach that avoids many common simplifications that have previously hampered realistic computational studies of molecular interactions between adjacent cellular membranes.
\section{Discussion}\label{discussion}
The challenges of choosing the right spatial resolution for computational representations of complex geometries with inhomogeneously distributed reactants becomes particularly difficult to address when stochasticity and proximity effects have to be considered. 
Whereas, in principle, spatial extensions of the SSA can be designed with such highly resolved spatial discretizations that effects such as geminate recombination can be reproduced correctly, 
the computational cost may become prohibitively high. Moreover, finding the appropriate resolution and/or adjusting reaction and diffusion events to the system one wishes to simulate can be non-trivial. 
Here, we introduced an approach that avoids these issues by abandoning the need for spatial discretization altogether, thereby providing a framework for grid-free stochastic single-particle based simulations of reaction-diffusion processes operating in non-trivial spatial contexts.
Our modeling approach circumvents the difficult problem of efficiently constructing appropriate explicit parameter maps for general surfaces. It offers the additional advantage of readily allowing for complex geometries while offering a highly efficient method for the computation of the local geometry required for the execution of the stochastic solver. Finally, the framework introduced here does not hinge on details of the stochastic solver but is compatible with a variety of different particle-based stochastic simulation methods.

We tested the validity of our new algorithm by modeling reversible and irreversible associations among molecules on curved surfaces and found that it provides excellent accuracy at the two-particle Green’s function level for the case of diffusion-influenced reactions on the surface of a sphere. We applied the algorithm also to a simple previously published model of the early phase of T-cell receptor activation and obtained results consistent with Ref.~\cite{fernandes2019cell} that had used a different simulation approach.

One should note that our algorithm still makes several simplifying assumptions. A case in point is the depiction of complexes, such as CD45, as rigid sticks of a certain height that are excluded in an all-or-nothing fashion from cell-cell contact regions with an incompatible gap distance. Another is the characterization of the diffusive motion of intermembrane complexes, such as bound TCR:pMHC complexes, in terms of a single diffusion coefficient parameter, despite the special spatial constraints that only exist for diffusing complexes spanning the gap between two cell membranes at contact. Work to address these simplifications is ongoing. 
\section{Appendix}\label{appendix}
The time evolution of the Green's function $p_{\text{free}}(\theta, t\vert \theta_{0})$ that describes the diffusion of a molecule with diffusion constant $D$ on a sphere with radius $R$ is governed by the diffusion equation
\begin{equation}\label{diff_eq_sphere}
\partial_{t} \,p_{\text{free}}(\theta, t\vert \theta_{0}) = \frac{D}{R^{2}\sin\theta}\partial_{\theta}\sin\theta\partial_{\theta} \,p_{\text{free}}(\theta, t\vert \theta_{0})
\end{equation}
subject to the initial condition
\begin{equation}\label{ic}
p_{\text{free}}(\theta, t=0\vert \theta_{0})=\frac{\delta(\theta - \theta_{0})}{2\pi R^{2} \sin\theta},
\end{equation}
where $\theta$ denotes the polar angle of standard spherical coordinates.
Here and in the following we will use the notation $\partial_{t} \equiv \partial/\partial t$, $\partial_{\theta} \equiv \partial/\partial \theta$ and so forth.
Grebenkov derived an exact expression for the survival probability of an isolated pair of molecules $A$, $B$ undergoing a diffusion-influenced reversible reaction on the surface of a sphere \cite{grebenkov2019reversible}.
The analytical form of the Green's function used in Sec.~\ref{numeric_GF_app} to probe the accuracy of the developed algorithm can be readily obtained from Grebenkov's results \cite{grebenkov2019reversible}.

As discussed in Sec.~\ref{bimol_reactions_mem}, the SCK approach introduces chemical reactions by imposing boundary conditions at the encounter distance $a_{\text{enc}}$. On a sphere, the to Eq.~(\ref{backreaction_bc}) corresponding backreaction boundary condition that describes reversible reactions assumes the form
\begin{equation}
\label{backreaction_bc_sphere}
-2\pi a_{\text{enc}} \frac{D}{R}\partial_{\theta} p_{\text{rev}}(\theta, t \vert \theta_{0})\vert_{\theta=\theta_{a}} = \kappa_{a} p_{\text{rev}}(\theta_{a}, t \vert \theta_{0}) - \kappa_{d} [1-S_{\text{rev}}(t\vert \theta_{0})],
\end{equation}
The polar angle $\theta_{a}$ describes the position of molecule $A$ when being at contact with molecule $B$ and is given by 
\begin{equation}\label{theta_a_def}
\cos\theta_{a} = -\sqrt{1- (a_{\text{enc}}/R)^{2}}, 
\end{equation}
Finally, the survival probability is defined by
\begin{equation}
S_{\text{rev}}(t\vert \theta_{0}) = 2\pi R^{2}\int^{\pi}_{\theta_{a}} p_{\text{rev}}(\theta, t \vert \theta_{0}) \sin\theta d\theta.
\end{equation}
The solution $p_{\text{rev}}(\theta, t \vert \theta_{0})$ of the initial-boundary value problem specified by Eqs.~(\ref{diff_eq_sphere}), (\ref{ic}), and (\ref{backreaction_bc_sphere}) takes the form \cite{grebenkov2019reversible}
\begin{equation}\label{p_abs_sphere}
p_{\text{rev}}(\theta, t \vert \theta_{0}) = \frac{1}{2\pi R^{2}}\sum_{n=0}^{\infty}\frac{b^{2}_{n}}{\partial_{s}f_{n}(s_{n})}\exp\bigg(-\frac{Dt \nu_{n}(\nu_{n}+1)}{R^{2}}\bigg)P_{\nu_{n}}(\cos \theta)P_{\nu_{n}}(\cos \theta_{0}).
\end{equation} 
In contrast to the free diffusion case \cite{brillinger1997particle}, $P_{\nu_{n}}$ now denotes Legendre functions of fractional order that are defined in terms of Gauss' hypergeometric function $_{2}F(a,b;c;x)$ \cite{abramowitz1964handbook}
\begin{equation}
P_{\nu}(x) := \, _{2}F(-\nu, \nu+1; 1; (1-x)/2)).
\end{equation}
To numerically evaluate $p_{\text{rev}}(\theta, t \vert \theta_{0})$, first one has to determine the $\nu_{n}$, which may be obtained as the solutions of \cite{grebenkov2019reversible}
\begin{equation}\label{roots_v}
[\bar{\kappa}_{d} - \nu(\nu+1)]P^{\prime}_{\nu}(\cos(\theta_{a})) = -q\nu(\nu+1)P_{\nu}(\cos(\theta_{a})),
\end{equation}
where the dimensionless quantities $\bar{\kappa}_{d}$ and $q$ are defined as
\begin{equation}
\bar{\kappa}_{d} = \kappa_{d}D/R^{2}, \qquad q = \kappa_{a} R^{2}/(2\pi a_{\text{enc}}^{2} D).
\end{equation}
Then, knowledge of the $\nu_{n}$ enables one to calculate the corresponding $b_{n}$ terms according to \cite{grebenkov2019reversible}
\begin{equation}
b_{n} = \bigg(\int^{1}_{a} [P_{\nu_{n}}(x)]^{2}\,dx\bigg)^{-1/2}.
\end{equation}

Finally, one has to calculate the $\partial_{s}f_{n}(s_{n})$ terms that appear in Eq.~(\ref{p_abs_sphere}). Here, $s_{n}$ refers to the poles of the Laplace-transformed GF 
$\tilde{p}_{\text{rev}}(\theta, s \vert \theta_{0})=\int^{\infty}_{0} p_{\text{rev}}(\theta, t \vert \theta_{0}) e^{-st}\,dt$. The poles can be obtained as the roots of the function 
\begin{equation}\label{f_n}
f_{n}(s)=s+\frac{D}{R^{2}}\nu_{n}(\nu_{n}+1).
\end{equation}
It follows that the derivative of $f_{n}$ is
\begin{equation}\label{d_f_n}
\partial_{s}f_{n}(s_{n})=1+\frac{D}{R^{2}}(2\nu_{n}+1)\partial_{s}\nu_{n}(s_{n}),
\end{equation}
where \cite{grebenkov2019reversible}
\begin{equation}\label{d_v_n}
\partial_{s}\nu_{n}(s_{n})=\frac{\bar{\kappa}\kappa_{d}P_{\nu_{n}}(\cos\theta_{a})}{\sin\theta_{a}(s_{n}+\kappa_{d})^{2}}\bigg(\partial_{\nu}P_{\nu}(\cos\theta_{a}) - \frac{qs_{n}}{s_{n}+\kappa_{d}}\partial_{\nu}P^{\prime}_{\nu}(\cos\theta_{a})\bigg)^{-1}\bigg\vert_{\nu=\nu_{n}}
\end{equation}

The calculation of the roots $\nu_{n}$ and the $\partial_{s}f_{n}(s_{n})$ terms involving various derivatives of the Legendre functions (Eqs.~(\ref{roots_v}), (\ref{d_f_n}), (\ref{d_v_n})) is facilitated by the following identities \cite{grebenkov2019reversible}
\begin{eqnarray}
P_{\nu}^{\prime}(x) &=& (\nu+1)\frac{P_{\nu+1}(x) - x P_{\nu}(x)}{x^{2}-1}, \\
\partial_{\nu}P_{\nu}(\cos\theta) &=& -\frac{\sqrt{2}}{\pi} \int^{\theta}_{0} d\alpha \frac{\sin[(\nu+1/2) \alpha]}{\sqrt{\cos\alpha-\cos\theta}}, \\
\partial_{\nu} P^{\prime}_{\nu}(\cos\theta_{a})&=&\frac{P_{\nu}(\cos\theta_{a})}{\nu+1} +  \nonumber\\
&&\frac{\nu+1}{\cos^{2}\theta_{a} - 1}[\partial_{\nu} P_{\nu+1}(\cos\theta_{a}) - \cos\theta_{a}\partial_{\nu} P_{\nu}(\cos\theta_{a})].
\end{eqnarray}

In the limit of irreversible diffusion-influenced ($\kappa_{d}=0$) and -controlled reactions ($\kappa_{a}=\infty$), the numerical evaluation of $p_{\text{rev}}(\theta, t \vert \theta_{0})$ simplifies. In particular, Eq.~(\ref{roots_v}) reduces to
\begin{equation}
P^{\prime}_{\nu}(\cos(\theta_{a})) = -qP_{\nu}(\cos(\theta_{a}))
\end{equation}
and
\begin{equation}
P_{\nu}(\cos(\theta_{a})) = 0,
\end{equation}
for $p_{\text{rad}}(\theta, t \vert \theta_{0})$ and $p_{\text{abs}}(\theta, t \vert \theta_{0})$, respectively. Furthermore, for both limiting cases, all $\partial_{s}f_{n}(s_{n})$ terms assume the value unity and in the case of of a perfectly absorbing sink, Eq.~(\ref{p_abs_sphere}) reduces to the expression for $p_{\text{abs}}(\theta, t \vert \theta_{0})$ obtained in Ref.~\cite{chao1981localization}.

To create Figs.~\ref{fig:p_sphere_abs}, \ref{fig:p_sphere_rad}, and \ref{fig:p_sphere_rev}, we calculated the first twenty terms of the expressions for $p_{\text{abs}}(\theta, t \vert \theta_{0})$, $p_{\text{rad}}(\theta, t \vert \theta_{0})$ and $p_{\text{rev}}(\theta, t \vert \theta_{0})$ (Eq.~(\ref{p_abs_sphere})), respectively.
\section*{Acknowledgments}
This research was supported by the Intramural Research Program of the NIH, NIAID. 
\bibliographystyle{plain} 


\begin{thebibliography}{10}

\bibitem{abramowitz1964handbook}
M.~Abramowitz and I.A. Stegun.
\newblock {\em Handbook of Mathematical Functions with Formulas, Graphs, and
  Mathematical Tables}.
\newblock Dover, New York, 1965.

\bibitem{agbanusi2014comparison}
Ikemefuna~C Agbanusi and Samuel~A Isaacson.
\newblock A comparison of bimolecular reaction models for stochastic
  reaction--diffusion systems.
\newblock {\em Bulletin of mathematical biology}, 76(4):922--946, 2014.

\bibitem{agmon1984diffusion}
Noam Agmon.
\newblock Diffusion with back reaction.
\newblock {\em The Journal of chemical physics}, 81(6):2811--2817, 1984.

\bibitem{agmon1990theory}
Noam Agmon and Attila Szabo.
\newblock Theory of reversible diffusion-influenced reactions.
\newblock {\em The Journal of Chemical Physics}, 92(9):5270--5284, 1990.

\bibitem{akenine2018real}
Tomas Akenine-Moller, Eric Haines, Naty Hoffman, et~al.
\newblock {\em Real-time rendering}.
\newblock AK Peters/CRC Press, 4th edition, 2018.

\bibitem{andrews2017smoldyn}
Steven~S Andrews.
\newblock Smoldyn: particle-based simulation with rule-based modeling, improved
  molecular interaction and a library interface.
\newblock {\em Bioinformatics}, 33(5):710--717, 2017.

\bibitem{andrews2018particle}
Steven~S Andrews.
\newblock Particle-based stochastic simulators.
\newblock {\em Encyclopedia of Computational Neuroscience}, 10:978--1, 2018.

\bibitem{angermann2012computational}
Bastian~R Angermann, Frederick Klauschen, Alex~D Garcia, Thorsten Prustel,
  Fengkai Zhang, Ronald~N Germain, and Martin Meier-Schellersheim.
\newblock Computational modeling of cellular signaling processes embedded into
  dynamic spatial contexts.
\newblock {\em Nature methods}, 9(3):283--289, 2012.

\bibitem{arjunan2013cell}
Satya Nanda~Vel Arjunan, Pawan~K Dhar, and Masaru Tomita.
\newblock {\em E-cell system: basic concepts and applications}.
\newblock Springer, 2013.

\bibitem{barenbrug2002accurate}
Theo~MAOM Barenbrug, EAJF Peters, and Jay~D Schieber.
\newblock Accurate method for the brownian dynamics simulation of spherical
  particles with hard-body interactions.
\newblock {\em The Journal of chemical physics}, 117(20):9202--9214, 2002.

\bibitem{belardi2020cell}
Brian Belardi, Sungmin Son, James~H Felce, Michael~L Dustin, and Daniel~A
  Fletcher.
\newblock Cell--cell interfaces as specialized compartments directing cell
  function.
\newblock {\em Nature Reviews Molecular Cell Biology}, 21(12):750--764, 2020.

\bibitem{blinn1982generalization}
James~F Blinn.
\newblock A generalization of algebraic surface drawing.
\newblock {\em ACM transactions on graphics (TOG)}, 1(3):235--256, 1982.

\bibitem{blinov2017compartmental}
Michael~L Blinov, James~C Schaff, Dan Vasilescu, Ion~I Moraru, Judy~E Bloom,
  and Leslie~M Loew.
\newblock Compartmental and spatial rule-based modeling with virtual cell.
\newblock {\em Biophysical journal}, 113(7):1365--1372, 2017.

\bibitem{brillinger1997particle}
David~R Brillinger.
\newblock A particle migrating randomly on a sphere.
\newblock {\em Journal of Theoretical Probability}, 10:429--443, 1997.

\bibitem{chang2016initiation}
Veronica~T Chang, Ricardo~A Fernandes, Kristina~A Ganzinger, Steven~F Lee,
  Christian Siebold, James McColl, Peter J{\"o}nsson, Matthieu Palayret, Karl
  Harlos, Charlotte~H Coles, et~al.
\newblock Initiation of t cell signaling by cd45 segregation at'close
  contacts'.
\newblock {\em Nature immunology}, 17(5):574--582, 2016.

\bibitem{chao1981localization}
NM~Chao, SH~Young, and MM~Poo.
\newblock Localization of cell membrane components by surface diffusion into a"
  trap".
\newblock {\em Biophysical Journal}, 36(1):139--153, 1981.

\bibitem{chen2021trapping}
Kevin~Y Chen, Edward Jenkins, Markus K{\"o}rbel, Aleks Ponjavic, Anna~H
  Lippert, Ana~Mafalda Santos, Nicole Ashman, Caitlin O'Brien-Ball, Jemma
  McBride, David Klenerman, et~al.
\newblock Trapping or slowing the diffusion of t cell receptors at close
  contacts initiates t cell signaling.
\newblock {\em Proceedings of the National Academy of Sciences},
  118(39):e2024250118, 2021.

\bibitem{collins1949diffusion}
Frank~C Collins and George~E Kimball.
\newblock Diffusion-controlled reaction rates.
\newblock {\em Journal of colloid science}, 4(4):425--437, 1949.

\bibitem{cordoba2013large}
Shaun-Paul Cordoba, Kaushik Choudhuri, Hao Zhang, Marcus Bridge, Alp~Bugra
  Basat, Michael~L Dustin, and P~Anton van~der Merwe.
\newblock The large ectodomains of cd45 and cd148 regulate their segregation
  from and inhibition of ligated t-cell receptor.
\newblock {\em Blood, The Journal of the American Society of Hematology},
  121(21):4295--4302, 2013.

\bibitem{davis2006kinetic}
Simon~J Davis and P~Anton Van Der~Merwe.
\newblock The kinetic-segregation model: Tcr triggering and beyond.
\newblock {\em Nature immunology}, 7(8):803--809, 2006.

\bibitem{de2015survey}
Bruno~Rodrigues De~Ara{\'u}jo, Daniel~S Lopes, Pauline Jepp, Joaquim~A Jorge,
  and Brian Wyvill.
\newblock A survey on implicit surface polygonization.
\newblock {\em ACM Computing Surveys (CSUR)}, 47(4):1--39, 2015.

\bibitem{delon2000information}
J{\'e}r{\^o}me Delon and Ronald~N Germain.
\newblock Information transfer at the immunological synapse.
\newblock {\em Current Biology}, 10(24):R923--R933, 2000.

\bibitem{doi1976second}
Masao Doi.
\newblock Second quantization representation for classical many-particle
  system.
\newblock {\em Journal of Physics A: Mathematical and General}, 9(9):1465,
  1976.

\bibitem{doi1976stochastic}
Masao Doi.
\newblock Stochastic theory of diffusion-controlled reaction.
\newblock {\em Journal of Physics A: Mathematical and General}, 9(9):1479,
  1976.

\bibitem{drawert2016stochastic}
Brian Drawert, Andreas Hellander, Ben Bales, Debjani Banerjee, Giovanni
  Bellesia, Bernie~J Daigle~Jr, Geoffrey Douglas, Mengyuan Gu, Anand Gupta,
  Stefan Hellander, et~al.
\newblock Stochastic simulation service: bridging the gap between the
  computational expert and the biologist.
\newblock {\em PLoS computational biology}, 12(12):e1005220, 2016.

\bibitem{edelstein1993brownian}
Arieh~L Edelstein and Noam Agmon.
\newblock Brownian dynamics simulations of reversible reactions in one
  dimension.
\newblock {\em The Journal of chemical physics}, 99(7):5396--5404, 1993.

\bibitem{edelstein1997brownian}
Arieh~L Edelstein and Noam Agmon.
\newblock Brownian simulation of many-particle binding to a reversible receptor
  array.
\newblock {\em Journal of Computational Physics}, 132(2):260--275, 1997.

\bibitem{fange2010stochastic}
David Fange, Otto~G Berg, Paul Sj{\"o}berg, and Johan Elf.
\newblock Stochastic reaction-diffusion kinetics in the microscopic limit.
\newblock {\em Proceedings of the National Academy of Sciences},
  107(46):19820--19825, 2010.

\bibitem{fernandes2019cell}
Ricardo~A Fernandes, Kristina~A Ganzinger, Justin~C Tzou, Peter J{\"o}nsson,
  Steven~F Lee, Matthieu Palayret, Ana~Mafalda Santos, Alexander~R Carr, Aleks
  Ponjavic, Veronica~T Chang, et~al.
\newblock A cell topography-based mechanism for ligand discrimination by the t
  cell receptor.
\newblock {\em Proceedings of the National Academy of Sciences},
  116(28):14002--14010, 2019.

\bibitem{frisvad2012building}
Jeppe~Revall Frisvad.
\newblock Building an orthonormal basis from a 3d unit vector without
  normalization.
\newblock {\em Journal of Graphics Tools}, 16(3):151--159, 2012.

\bibitem{gardiner2009stochastic}
Crispin Gardiner.
\newblock {\em Stochastic methods}, volume~4.
\newblock Springer Berlin, 2009.

\bibitem{gardiner1976correlations}
CW~Gardiner, KJ~McNeil, DF~Walls, and IS~Matheson.
\newblock Correlations in stochastic theories of chemical reactions.
\newblock {\em Journal of Statistical Physics}, 14:307--331, 1976.

\bibitem{gillespie1976general}
Daniel~T Gillespie.
\newblock A general method for numerically simulating the stochastic time
  evolution of coupled chemical reactions.
\newblock {\em Journal of computational physics}, 22(4):403--434, 1976.

\bibitem{gillespie1977exact}
Daniel~T Gillespie.
\newblock Exact stochastic simulation of coupled chemical reactions.
\newblock {\em The journal of physical chemistry}, 81(25):2340--2361, 1977.

\bibitem{gomes2009implicit}
Abel~JP Gomes, Irina Voiculescu, Joaquim Jorge, Brian Wyvill, and Callum
  Galbraith.
\newblock {\em Implicit curves and surfaces: mathematics, data structures and
  algorithms}.
\newblock Springer, 2009.

\bibitem{gosele1984reaction}
Ulrich~M G{\"o}sele.
\newblock Reaction kinetics and diffusion in condensed matter.
\newblock {\em Progress in reaction kinetics}, 13(2), 1984.

\bibitem{grebenkov2019reversible}
Denis~S Grebenkov.
\newblock Reversible reactions controlled by surface diffusion on a sphere.
\newblock {\em The Journal of Chemical Physics}, 151(15), 2019.

\bibitem{gruenert2010rule}
Gerd Gruenert, Bashar Ibrahim, Thorsten Lenser, Maiko Lohel, Thomas Hinze, and
  Peter Dittrich.
\newblock Rule-based spatial modeling with diffusing, geometrically constrained
  molecules.
\newblock {\em BMC bioinformatics}, 11:1--14, 2010.

\bibitem{hepburn2012steps}
Iain Hepburn, Weiliang Chen, Stefan Wils, and Erik De~Schutter.
\newblock Steps: efficient simulation of stochastic reaction--diffusion models
  in realistic morphologies.
\newblock {\em BMC systems biology}, 6(1):1--19, 2012.

\bibitem{hoffmann2019readdy}
Moritz Hoffmann, Christoph Fr{\"o}hner, and Frank No{\'e}.
\newblock Readdy 2: Fast and flexible software framework for
  interacting-particle reaction dynamics.
\newblock {\em PLoS computational biology}, 15(2):e1006830, 2019.

\bibitem{holyst1999diffusion}
R~Ho{\l}yst, D~Plewczy{\'n}ski, A~Aksimentiev, and K~Burdzy.
\newblock Diffusion on curved, periodic surfaces.
\newblock {\em Physical Review E}, 60(1):302, 1999.

\bibitem{hughes1999building}
John~F Hughes and Tomas Moller.
\newblock Building an orthonormal basis from a unit vector.
\newblock {\em Journal of Graphics Tools}, 4(4):33--35, 1999.

\bibitem{husar2022mcell4}
Adam Husar, Mariam Ordyan, Guadalupe~C Garcia, Joel~G Yancey, Ali~S Saglam,
  James~R Faeder, Thomas~M Bartol, Mary~B Kennedy, and Terrence~J Sejnowski.
\newblock Mcell4 with bionetgen: A monte carlo simulator of rule-based
  reaction-diffusion systems with python interface.
\newblock {\em bioRxiv}, pages 2022--05, 2022.

\bibitem{isaacson2013convergent}
Samuel~A Isaacson.
\newblock A convergent reaction-diffusion master equation.
\newblock {\em The Journal of chemical physics}, 139(5), 2013.

\bibitem{isaacson2018unstructured}
Samuel~A Isaacson and Ying Zhang.
\newblock An unstructured mesh convergent reaction--diffusion master equation
  for reversible reactions.
\newblock {\em Journal of Computational Physics}, 374:954--983, 2018.

\bibitem{jenkins2023antigen}
Edward Jenkins, Markus K{\"o}rbel, Caitlin O’Brien-Ball, James McColl,
  Kevin~Y Chen, Mateusz Kotowski, Jane Humphrey, Anna~H Lippert, Heather
  Brouwer, Ana~Mafalda Santos, et~al.
\newblock Antigen discrimination by t cells relies on size-constrained
  microvillar contact.
\newblock {\em Nature Communications}, 14(1):1611, 2023.

\bibitem{johnson2021quantifying}
Margaret~E Johnson, Athena Chen, James~R Faeder, Philipp Henning, Ion~I Moraru,
  Martin Meier-Schellersheim, Robert~F Murphy, Thorsten Pr{\"u}stel, Julie~A
  Theriot, and Adelinde~M Uhrmacher.
\newblock Quantifying the roles of space and stochasticity in computer
  simulations for cell biology and cellular biochemistry.
\newblock {\em Molecular Biology of the Cell}, 32(2):186--210, 2021.

\bibitem{johnson2014free}
Margaret~E Johnson and Gerhard Hummer.
\newblock Free-propagator reweighting integrator for single-particle dynamics
  in reaction-diffusion models of heterogeneous protein-protein interaction
  systems.
\newblock {\em Physical Review X}, 4(3):031037, 2014.

\bibitem{kim1999exact}
Hyojoon Kim and Kook~Joe Shin.
\newblock Exact solution of the reversible diffusion-influenced reaction for an
  isolated pair in three dimensions.
\newblock {\em Physical review letters}, 82(7):1578, 1999.

\bibitem{kim1999dynamic}
Hyojoon Kim, Mino Yang, and Kook~Joe Shin.
\newblock Dynamic correlation effect in reversible diffusion-influenced
  reactions: Brownian dynamics simulation in three dimensions.
\newblock {\em The Journal of chemical physics}, 111(3):1068--1075, 1999.

\bibitem{manz2010spatial}
Boryana~N Manz and Jay~T Groves.
\newblock Spatial organization and signal transduction at intercellular
  junctions.
\newblock {\em Nature Reviews Molecular Cell Biology}, 11(5):342--352, 2010.

\bibitem{michalski2016springsalad}
Paul~J Michalski and Leslie~M Loew.
\newblock Springsalad: a spatial, particle-based biochemical simulation
  platform with excluded volume.
\newblock {\em Biophysical journal}, 110(3):523--529, 2016.

\bibitem{novak2007diffusion}
Igor~L Novak, Fei Gao, Yung-Sze Choi, Diana Resasco, James~C Schaff, and
  Boris~M Slepchenko.
\newblock Diffusion on a curved surface coupled to diffusion in the volume:
  Application to cell biology.
\newblock {\em Journal of computational physics}, 226(2):1271--1290, 2007.

\bibitem{oppelstrup2009first}
Tomas Oppelstrup, Vasily~V Bulatov, Aleksandar Donev, Malvin~H Kalos, George~H
  Gilmer, and Babak Sadigh.
\newblock First-passage kinetic monte carlo method.
\newblock {\em Physical Review E}, 80(6):066701, 2009.

\bibitem{prustel2012exact}
Thorsten Pr{\"u}stel and Martin Meier-Schellersheim.
\newblock Exact green's function of the reversible diffusion-influenced
  reaction for an isolated pair in two dimensions.
\newblock {\em The Journal of chemical physics}, 137(5), 2012.

\bibitem{prustel2017unified}
Thorsten Pr{\"u}stel and Martin Meier-Schellersheim.
\newblock Unified path integral approach to theories of diffusion-influenced
  reactions.
\newblock {\em Physical Review E}, 96(2):022151, 2017.

\bibitem{prustel2020stochastic}
Thorsten Pr{\"u}stel and Martin Meier-Schellersheim.
\newblock Stochastic single-particle based simulations of cellular signaling
  embedded into computational models of cellular morphology.
\newblock {\em arXiv preprint arXiv:2003.13634}, 2020.

\bibitem{prustel2021space}
Thorsten Pr{\"u}stel and Martin Meier-Schellersheim.
\newblock Space--time histories approach to fast stochastic simulation of
  bimolecular reactions.
\newblock {\em The Journal of Chemical Physics}, 154(16), 2021.

\bibitem{rao2002control}
Christopher~V Rao, Denise~M Wolf, and Adam~P Arkin.
\newblock Control, exploitation and tolerance of intracellular noise.
\newblock {\em Nature}, 420(6912):231--237, 2002.

\bibitem{rice1985diffusion}
Stephen~A Rice.
\newblock {\em Diffusion-limited reactions}.
\newblock Elsevier, New York, 1985.

\bibitem{roberts2013lattice}
Elijah Roberts, John~E Stone, and Zaida Luthey-Schulten.
\newblock Lattice microbes: High-performance stochastic simulation method for
  the reaction-diffusion master equation.
\newblock {\em Journal of computational chemistry}, 34(3):245--255, 2013.

\bibitem{shoup1982role}
David Shoup and Attila Szabo.
\newblock Role of diffusion in ligand binding to macromolecules and cell-bound
  receptors.
\newblock {\em Biophysical Journal}, 40(1):33--39, 1982.

\bibitem{smoluchowski1917mathematical}
MV~Smoluchowski.
\newblock Mathematical theory of the kinetics of the coagulation of colloidal
  solutions.
\newblock {\em Z. Phys. Chem.}, 92:129--168, 1917.

\bibitem{sneddon2011efficient}
Michael~W Sneddon, James~R Faeder, and Thierry Emonet.
\newblock Efficient modeling, simulation and coarse-graining of biological
  complexity with nfsim.
\newblock {\em Nature methods}, 8(2):177--183, 2011.

\bibitem{sokolowski2019egfrd}
Thomas~R Sokolowski, Joris Paijmans, Laurens Bossen, Thomas Miedema, Martijn
  Wehrens, Nils~B Becker, Kazunari Kaizu, Koichi Takahashi, Marileen Dogterom,
  and Pieter~Rein Ten~Wolde.
\newblock egfrd in all dimensions.
\newblock {\em The Journal of chemical physics}, 150(5), 2019.

\bibitem{takahashi2010spatio}
Koichi Takahashi, Sorin T{\u{a}}nase-Nicola, and Pieter~Rein Ten~Wolde.
\newblock Spatio-temporal correlations can drastically change the response of a
  mapk pathway.
\newblock {\em Proceedings of the National Academy of Sciences},
  107(6):2473--2478, 2010.

\bibitem{tolle2010meredys}
Dominic~P Tolle and Nicolas Le~Nov{\`e}re.
\newblock Meredys, a multi-compartment reaction-diffusion simulator using
  multistate realistic molecular complexes.
\newblock {\em BMC systems biology}, 4(1):1--11, 2010.

\bibitem{van1992stochastic}
Nicolaas~Godfried Van~Kampen.
\newblock {\em Stochastic processes in physics and chemistry}, volume~1.
\newblock Elsevier, 1992.

\bibitem{van2005simulating}
Jeroen~S van Zon and Pieter~Rein Ten~Wolde.
\newblock Simulating biochemical networks at the particle level and in time and
  space: Green’s function reaction dynamics.
\newblock {\em Physical review letters}, 94(12):128103, 2005.

\bibitem{varga2020nerdss}
Matthew~J Varga, Yiben Fu, Spencer Loggia, Osman~N Yogurtcu, and Margaret~E
  Johnson.
\newblock Nerdss: a nonequilibrium simulator for multibody self-assembly at the
  cellular scale.
\newblock {\em Biophysical Journal}, 118(12):3026--3040, 2020.

\bibitem{weaver1983diffusion}
DL~Weaver.
\newblock Diffusion-mediated localization on membrane surfaces.
\newblock {\em Biophysical Journal}, 41(1):81, 1983.

\bibitem{yogurtcu2015theory}
Osman~N Yogurtcu and Margaret~E Johnson.
\newblock Theory of bi-molecular association dynamics in 2d for accurate model
  and experimental parameterization of binding rates.
\newblock {\em The Journal of Chemical Physics}, 143(8), 2015.

\end{thebibliography}

\end{document}